%% file: finalfits.tex
\pdfoutput=1
\documentclass[aps,prd,amsmath,floats,floatfix, twocolumn, superscriptaddress,nofootinbib,showpacs]{revtex4-1}
\input{preamble}

\begin{document}

\title{\mbox{High-accuracy mass, spin, and recoil predictions of generic
black-hole merger remnants}}

\input{authors}

\date{\today}

\input{abstract.tex}

\maketitle

\input{body-content}

\input{surfin_letter.bbl}
\clearpage

\begin{center}
  \textbf{Supplemental Material} \\
\end{center}

\input{supplement-content}

\input{surfin_supp.bbl}
\end{document}

%% file: preamble.tex
\usepackage[T1]{fontenc}
\usepackage[utf8]{inputenc}
\usepackage{lmodern}

\usepackage{verbatim}

\usepackage[dvipsnames, usenames]{xcolor}
\definecolor{linkcolor}{rgb}{0.0,0.3,0.5}
\usepackage[hypertexnames=false, unicode, colorlinks=true, linkcolor=linkcolor, citecolor=linkcolor, filecolor=linkcolor,urlcolor=linkcolor, pdfusetitle]{hyperref}
\usepackage[all]{hypcap}
\usepackage{graphicx}
\usepackage{xspace}
\usepackage{amssymb}
\usepackage[normalem]{ulem}
\usepackage{bm}

\usepackage{microtype}

\graphicspath{%
  {figs/}%
}

\DeclareMathAlphabet{\mathpzc}{OT1}{pzc}{m}{it}

\newcommand{\roughly}{\mathchar"5218\relax\,}
\newcommand{\into}{\!\times\!\relax}

\newcommand{\chieff}{\chi_{\mathrm{eff}}}

\newcommand{\bchi}{\bm{\chi}}
\newcommand{\bv}{\bm{v}}

\newcommand{\ts}{\mathcal{TS}}
\newcommand{\ps}{\mathcal{PS}}

\newcommand{\PackageName}{\emph{surfinBH}\xspace}
\newcommand{\PrecessingModelName}{\emph{surfinBH7dq2}\xspace}
\newcommand{\AlignedModelName}{\emph{surfinBH3dq8}\xspace}

\def\apj{\rm{ApJ}}
\def\apjl{\rm{ApJ}}

\def\prd{\rm{PRD}}
\def\prl{\rm{PRL}}
\def\prx{\rm{PRX}}

\def\cqg{\rm{CQG}}

\def\lrr{\rm{LRR}}

\newcommand\prlsec[1]{\vspace{2mm}\noindent \emph{#1}--}

%% file: authors.tex
\newcommand\caltech{\affiliation{TAPIR 350-17, California Institute of Technology, 1200 E California Boulevard, Pasadena, CA 91125, USA}}
\newcommand\olemiss{\affiliation{Department of Physics and Astronomy,
The University of Mississippi, University, MS 38677, USA}} 

\author{Vijay Varma}
\email{vvarma@caltech.edu}
\caltech

\author{Davide Gerosa}
\thanks{Einstein Fellow}
\email{dgerosa@caltech.edu}
\caltech

\author{Leo C.~Stein}
\email{lcstein@olemiss.edu}
\caltech \olemiss

\author{Fran\c{c}ois~H\'{e}bert}
\email{fhebert@caltech.edu}
\caltech

\author{Hao Zhang}
\email{zhangphy@sas.upenn.edu}
\caltech
\affiliation{Department of Physics and Astronomy, University of Pennsylvania,
Philadelphia, PA 19104, USA}

\hypersetup{pdfauthor={Varma et al.}}

%% file: abstract.tex
\begin{abstract}
We present accurate fits for the remnant properties of generically precessing
binary black holes, trained on large banks of numerical-relativity simulations.
We use Gaussian process regression to interpolate the remnant mass, spin, and
recoil velocity in the 7-dimensional parameter space of precessing black-hole
binaries with mass ratios $q\leq2$, and spin magnitudes $\chi_1,\chi_2\leq0.8$.
For precessing systems, our errors in estimating the remnant mass, spin
magnitude, and kick magnitude are lower than those of existing fitting formulae
by at least an order of magnitude (improvement is also reported in the extrapolated region at high mass ratios and spins). In addition, we also model the remnant spin
and kick directions. Being trained directly on precessing simulations, our fits
are free from ambiguities regarding the initial frequency at which precessing
quantities are defined. We also construct a model for remnant properties of
aligned-spin systems with mass ratios $q\leq8$, and spin magnitudes
$\chi_1,\chi_2\leq0.8$.  As a byproduct, we also provide error estimates for
all fitted quantities, which can be consistently incorporated into current and
future gravitational-wave parameter-estimation analyses.  Our model(s) are made
publicly available through a fast and easy-to-use Python module called
\emph{surfinBH}.
\end{abstract}

%% file: body-content.tex
\begin{figure}[t]
  \centering
  \includegraphics[width=\columnwidth]{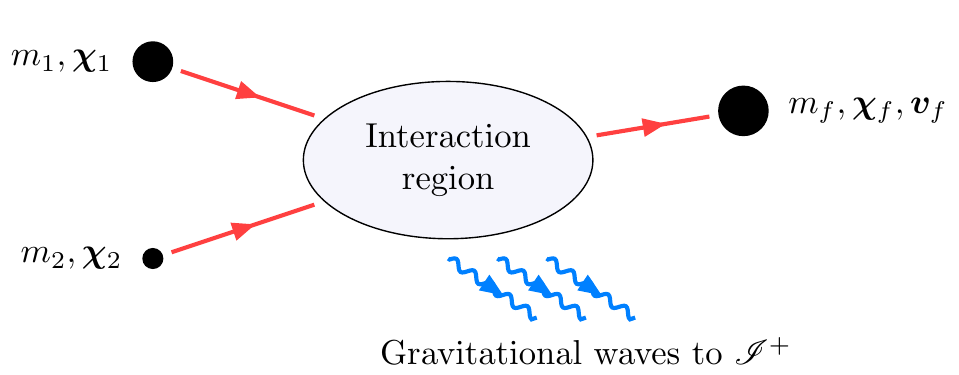}
  \caption{Quasi-circular binary BH merger problem viewed as a
    scattering process via a ``Feynman'' diagram.  Time flows to the
    right.  All quantities are well defined in the asymptotically flat
    region far from the interaction (merger).\vspace{-1em}   }
  \label{fig:interaction}
\end{figure}

\prlsec{Introduction} \label{sec:introduction}
As two black holes (BHs) come together and merge, they
emit copious gravitational waves (GWs) and leave behind a BH remnant.
The strong-field dynamics of this process are analytically intractable and must be
simulated using numerical relativity (NR).  However, from very far away,
the merger can be viewed as a scattering problem, depicted in
Fig.~\ref{fig:interaction}.  The complicated dynamics of the near zone
can be overlooked in favor of the gauge-invariant observables of the in-
and out-states: the initial BH masses and spins, the outgoing GWs, and
the final BH remnant.
This final BH is fully characterized by its mass, spin, and
recoil velocity; all additional complexities (``hair'') of the merging
binary are dissipated away in GWs~\cite{1968CMaPh...8..245I,1971PhRvL..26..331C,1996bhut.book.....H}.

All GW models designed to capture the entire inspiral-merger-ringdown (IMR)
signal from BH binary coalescences need to be calibrated to NR simulations
(e.g., \cite{2014PhRvL.113o1101H, 2016PhRvD..93d4007K, 2016PhRvD..93d4006H,
    2009PhRvD..79l4028B, 2017PhRvD..95d4028B, 2017PhRvD..95b4010B,
2014PhRvX...4c1006F, 2017PhRvD..95j4023B, 2017PhRvD..96b4058B}).  In
particular, the BH ringdown emission is crucially dependent on the properties
of the BH remnant --- properties obtained from NR simulations.  Accurate
modeling of the merger remnant is therefore vital for construction of accurate
IMR templates.

Besides waveform building, accurate knowledge of the remnant properties is also
instrumental to fulfill one of the greatest promises of GW astronomy: testing
Einstein's general relativity (GR) in its strong-field, highly dynamical
regime.  Current approaches to test the Kerr hypothesis attempt to measure the
properties of the inspiralling BHs from the low frequency part of the GW signal, then
use NR fits to predict the corresponding remnant mass and spin; this final-state
prediction
is compared to the properties inferred from the high frequency part of the GW signal
\cite{2018CQGra..35a4002G,2016PhRvL.116v1101A}.  Inaccuracies in remnant
models therefore directly propagate to the final fundamental-physics test.

The importance of building fits for the remnant properties was realized soon
after the NR breakthrough~\cite{2005PhRvL..95l1101P,2006PhRvL..96k1101C,2009PhRvD..79b4003S} and has been periodically revisited by several groups since then
\cite{2007PhRvD..76h4032H,
2007PhRvL..98w1102C,
2007PhRvL..98w1101G,
2007PhRvL..98i1101G,
2007ApJ...659L...5C,
2008PhRvD..78d4002R,
2008ApJ...674L..29R,
2008PhRvD..78h4030K,
2008PhRvD..78h1501T,
2008PhRvD..77d4028L,
2009ApJ...704L..40B,
2011PhRvD..84l4052P,
2012ApJ...758...63B,
2012PhRvD..85h4015L,
2013PhRvD..87h4027L,
2014PhRvD..90j4004H,
2015PhRvD..92b4022Z,
2016ApJ...825L..19H,
2016PhRvD..93l4066G,
2017PhRvD..95f4024J,
2017PhRvD..95b4037H,
2018PhRvD..97h4002H}.
There are two important shortcomings in all existing fitting formulae.
First, they enforce analytic ans\"atze (with NR-calibrated
coefficients) that are physically motivated, but lack a rigorous
mathematical justification.  Therefore, current fits can be prone to
systematic errors, especially in regions of parameter space where the
intuition used to design the formulae become less accurate.  Second, current
expressions for remnant mass and spins are calibrated on aligned-spin simulations and therefore fail to fully
capture the rich physics of precessing systems (but see e.g. \cite{2015PhRvD..92b4022Z} where a non-generic subspace of precessing configurations is considered). For example, current LIGO/Virgo
parameter-estimation pipelines \cite{2016PhRvX...6d1015A,2017PhRvL.118v1101A}
rely on ad-hoc
corrections to partially account for precession effects \cite{dccprec}. Aligned
fits applied to precessing systems are inevitably ambiguous, as the outcome
will depend on \emph{where} (in time, separation, or frequency) the spins are
defined and inserted into the fits (e.g., \cite{2010PhRvD..81h4054K}).

In this \emph{Letter} we tackle both these issues for the first time.  We
construct surrogate models that fit the remnant properties from a large sample
of generic, precessing, quasi-circular binary BH simulations performed with the Spectral Einstein
Code (SpEC)~\cite{2000PhRvD..62h4032K}. Surrogates are trained directly
against the NR simulations, using Gaussian process regression (GPR) without any
phenomenological ans\"atz, and achieve accuracies comparable to those of the
NR simulations themselves.
In their regime of validity, the models presented here are at least an order of
magnitude more accurate than previous fits.

In particular, we present two models:
\begin{enumerate}
\item \PrecessingModelName: a fit trained against precessing systems with mass
    ratios $q\leq2$ and dimensionless spin magnitudes $\chi_{1},\chi_{2}
    \leq0.8$.
\item \AlignedModelName: an aligned-spin model trained against systems with
    mass ratios up to $q\leq8$ and \mbox{(anti-)aligned} spin magnitudes
    $\chi_{1},\chi_{2}\leq0.8$.
\end{enumerate}
Both these models can be easily accessed using the publicly available Python
module \PackageName~\cite{surfinBH}, and are ready to be incorporated in both
waveform constructions and GW parameter-estimation studies.

\prlsec{Fitting procedure}
\label{sec:fit_procedure}
We construct fits for  the BH remnant mass $m_f$, spin vector $\bchi_f$, and
recoil kick vector $\bv_f$  as functions of the binary mass ratio $q$ and spin
vectors $\bchi_{1}, \bchi_{2}$. Our fits for \PrecessingModelName
(\AlignedModelName) map a 7- \mbox{(3-)}dimensional input parameter space to
a 7- \mbox{(4-)}dimensional output parameter space.  The fits are performed in the coorbital
frame at $t\!=\!-100M$, with $t\!=\!0$ at the peak of the total waveform
amplitude (cf. Ref.~\cite{2017PhRvD..96b4058B} for details). The coorbital frame is
defined such that the $z$-axis lies along the direction of the orbital angular
momentum, the $x$-axis runs from the smaller BH to the larger BH, and the
$y$-axis completes the triad.

All fits are performed using GPR~\cite{2006gpml.book.....R}; details are provided in the
supplemental material~\cite{surfsupplement}.  Notably, GPR naturally returns estimates of the errors
of the fitted quantities across the parameter space.

The values of spins, masses, and kicks used in the training process are extracted
directly from the NR simulations. We use the simulations
presented in Ref.~\cite{2017PhRvD..96b4058B} for \PrecessingModelName and those of
Ref.~\cite{Varma:2018hybsur} for \AlignedModelName.  Both spins and masses are
evaluated on apparent horizons \cite{2008PhRvD..78h4017L}; the dimensionful spin
$\bm{S}$ solves an eigenvalue problem for an approximate Killing vector, and the
mass is determined from the spin and area $A$ following the Christodoulou
relation $m^{2} = m_{\mathrm{irr}}^{2} + S^{2}/(4 m_{\mathrm{irr}}^{2})$, where
$m_{\mathrm{irr}}^{2} = A/16\pi$ is the irreducible mass.  The masses $m_{1,2}$ are
determined close to the beginning of the simulation at the ``relaxation time''
\cite{Catalog2018},
whereas the spins
$\bm{\chi}_{1,2}\equiv \bm{S}_{1,2}/m_{1,2}^{2}$ are measured at $t\!=\!-100M$. 
The remnant mass $m_{f}$ and spin $\bm{\chi}_{f}$ are 
determined long after ringdown, as detailed in~\cite{Catalog2018}.
All masses are in units of the total
mass $M=m_{1}+m_{2}$ at relaxation.  The remnant kick velocity is derived from conservation of momentum, $\bm{v}_{f} =
-\bm{P}^{\mathrm{rad}}/m_{f}$ \cite{2018PhRvD..97j4049G}.  The radiated momentum flux
$\bm{P}^{\mathrm{rad}}$ is integrated~\cite{2008GReGr..40.1705R} from the GWs
extrapolated to future null infinity \cite{2009PhRvD..80l4045B,Catalog2018}.
Before constructing the fits, $\bm{\chi}_f$ and $\bm{v}_f$ are transformed
into the coorbital frame at $t\!=\!-100M$.

Besides the GPR error estimate, we further address the accuracy of our
procedure using ``k-fold'' cross validations with $k\!=\!20$. First, we
randomly divide our training dataset into $k$ mutually exclusive sets. For each
set, we construct the fits using the data in the other $k-1$ sets and then test
the fits by evaluating them at the data points in the considered set.  We thus
obtain ``out-of-sample'' errors which conservatively indicate the
(in)accuracies of our fits.  We compare these errors against the intrinsic
error present in the NR waveforms, estimated by comparing the two highest
resolutions available.  We also compare the performance of our fits against
several existing fitting formulae for remnant mass, spin, and kick which we
denote as follows: HBMR (\cite{2016ApJ...825L..19H, 2012ApJ...758...63B} with
$n_M\!=\!n_J\!=\!3$), UIB \cite{2017PhRvD..95f4024J}, HL
\cite{2017PhRvD..95b4037H}, HLZ \cite{2014PhRvD..90j4004H}, and CLZM
(\cite{2007ApJ...659L...5C,2007PhRvL..98i1101G,2008PhRvD..77d4028L,2012PhRvD..85h4015L,2013PhRvD..87h4027L,2008PhRvD..77d4028L}
as summarized in \cite{2016PhRvD..93l4066G}).  To partially account for spin
precession, fits are corrected as described in Ref.~\cite{dccprec} and used in
current LIGO/Virgo analyses \cite{2016PhRvX...6d1015A,2017PhRvL.118v1101A}:
spins are evolved from relaxation to the Schwarzschild innermost stable
circular orbit, and final UIB and HL spins are post-processed adding the sum of
the in-plane spins in quadrature. We note these fitting formulae were
calibrated against different sets of simulations. Fitting methods, number of simulations, their quality, and
 their distribution in parameter space all contribute to the accuracy of the fits.

\begin{figure*}[p]
\centering
\includegraphics[scale=0.65]{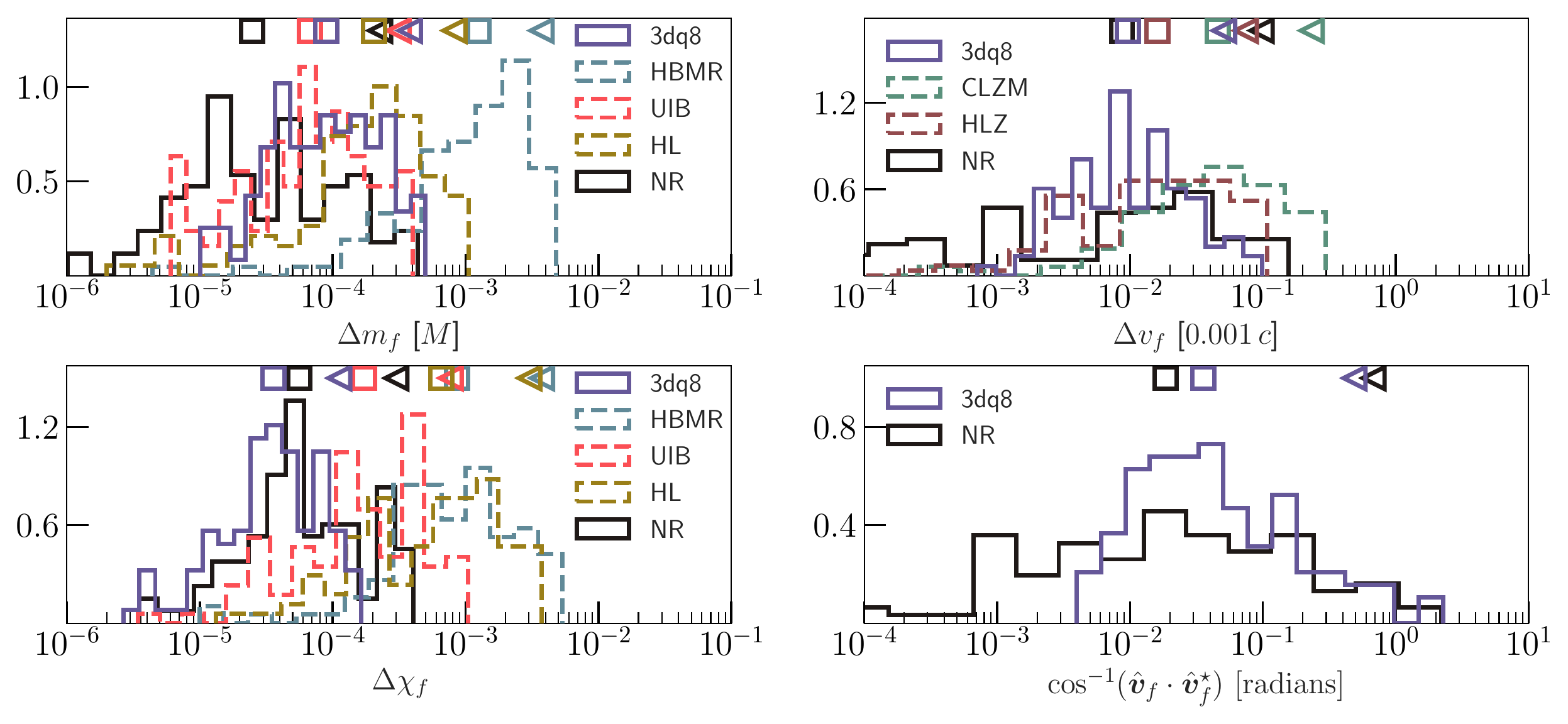}
\caption{Errors in predicting remnant mass, spin, kick magnitude, and
kick direction for non-precessing binary BHs with mass ratios $q\leq8$, and
spin magnitudes $\chi_{1},\chi_{2}\leq0.8$. The direction error is the
angle between the predicted vector and a fiducial vector, taken to be the
high-resolution NR case and indicated by a $^{\star}$. The square (triangle)
markers indicate median ($95^{\rm th}$
percentile) values.  Our model \AlignedModelName is referred to as 3dq8.  The
black histogram shows the NR resolution error while the dashed histograms show
errors for different existing fitting formulae.
}
\label{fig:aligned_spin_fits}
$\,$\\ $\,$\\
\capstart
\includegraphics[scale=0.65]{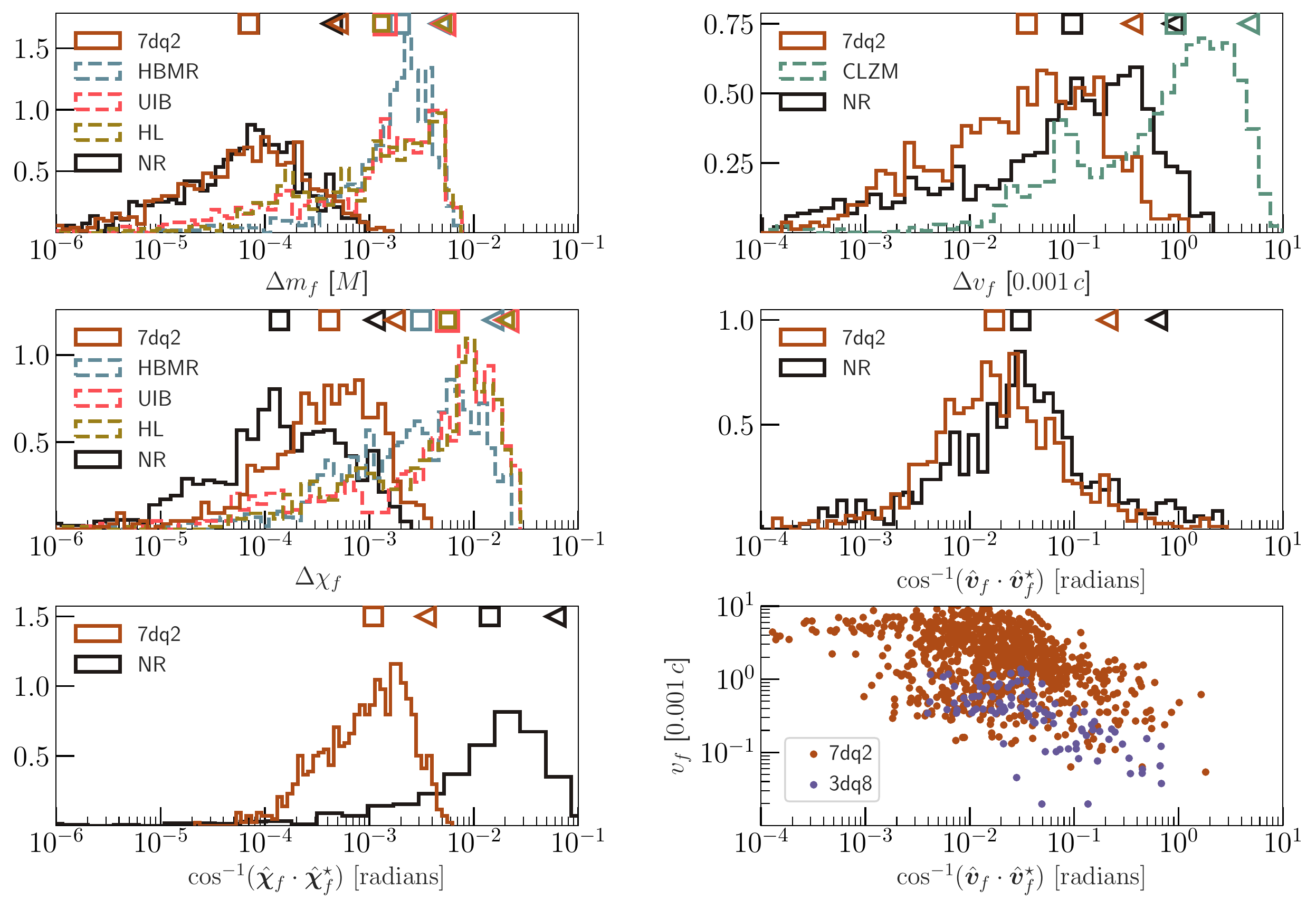}
\caption{Errors in predicting the remnant mass, spin magnitude, spin direction,
kick magnitude, and kick direction for precessing binary BHs with mass ratios $q\leq2$,
and spin magnitudes $\chi_{1},\chi_{2}\leq0.8$. Our model, \PrecessingModelName is referred to as 7dq2.  The black
histogram shows the NR resolution error while the dashed histograms show errors
for different existing fitting formulae. In the bottom-right panel we show the
distribution of kick magnitude vs. error in kick direction.
}
\label{fig:precessing_fits}
\end{figure*}

\prlsec{Aligned-spin model}
\label{sec:aligned_spin_fits}
We first present our fit \AlignedModelName, which is trained against 104
aligned-spin simulations~\cite{Varma:2018hybsur} with $q\leq 8$ and $-0.8\leq
\chi_{1z},\chi_{2z} \leq 0.8$. Symmetry implies that the kick lies in the
orbital plane while the final spin is orthogonal to it
\cite{2008PhRvL.100o1101B}. We therefore only fit for four quantities: $m_f$,
$\chi_{fz}$, $v_{fx}$, and $v_{fy}$.

Figure~\ref{fig:aligned_spin_fits} shows the out-of-sample errors of
\AlignedModelName. Our fits are as accurate as the NR simulations used in the
training process.  $95^{\rm th}$ percentile errors lie at $\Delta m_f \roughly
4\into10^{-4} M$, $\Delta\chi_f \roughly 10^{-4}$, and $\Delta v_f \roughly
5\into10^{-5} c$. The kick direction is predicted with an accuracy of $\roughly
0.5$ radians, which is the inherent accuracy of the NR simulations.
Our errors for the remnant mass and kick magnitude are comparable to the most
accurate existing fits. On the other hand, for the final spin, our procedure
outperforms all other formulae by at least a factor of\nobreak{} $5$.

\prlsec{Precessing model}
\label{sec:precessing_fits}
We now present \PrecessingModelName, a remnant model trained on 890
simulations~\cite{2017PhRvD..96b4058B} of generic, fully precessing BH binaries
with mass ratios $q\leq2$ and spin magnitudes $\chi_{1},\chi_{2} \leq 0.8$.
Out-of-sample errors are shown in Fig.~\ref{fig:precessing_fits}. $95^{\rm th}$
percentiles are $\roughly 5 \into 10^{-4} M$ for mass, $\roughly 2 \into
10^{-3}$ for spin magnitude, $\roughly 4 \into 10^{-3}$ radians for spin
direction, $\roughly 4 \into 10^{-4}\,c$ for kick magnitude, and $\roughly 0.2$
radians for kick direction.  As in the aligned-spin case above, our errors are
at the same level as the NR resolution error, thus showing that we are not
limited by our fitting procedure but rather by the quality of the training
dataset. Our fits appear to outperform the NR simulations when estimating the
spin direction, which suggests this quantity has not fully converged in the
NR\nobreak{} runs, and that the difference between the two highest resolution
simulations is an overestimate of the NR error in this quantity.

Figure \ref{fig:precessing_fits} shows that our procedure to predict remnant
mass, spin magnitude, and kick magnitude for precessing systems is more precise
than all existing fits by at least an order of magnitude. These
existing fits presented significantly lower errors when applied to aligned binaries (cf.
Fig.~\ref{fig:aligned_spin_fits}), which suggests that they fail to
fully capture precession effects despite the augmentation of
Ref.~\cite{dccprec}.  Some impact of precession effects on the final spin and recoil
is expected, since both of these quantities have been found to depend
strongly on the in-plane orientations of the spins of the merging BHs
\cite{2010PhRvD..81h4054K,2012PhRvD..85l4049B,2018PhRvD..97j4049G}.  More
surprisingly, we find that spin precession significantly affects the energy
radiated as well, which was expected to depend mostly on the aligned-spin
components via the orbital hang-up effect
\cite{2006PhRvD..74d1501C,2014PhRvD..89b1501L,2015CQGra..32j5009S}.

The largest errors in the kick direction can be of order $\roughly 1$ radian.
The bottom-right panel of Fig.~\ref{fig:precessing_fits} shows the joint
distribution of kick magnitude and kick direction error for both
\PrecessingModelName and \AlignedModelName, showing that errors are larger at
low kick magnitudes.  Our error in kick direction is below $\roughly 0.1$
radians whenever \mbox{$v_f\gtrsim 10^{-3} c$.}

\begin{figure}[]
\centering
\includegraphics[width=\columnwidth]{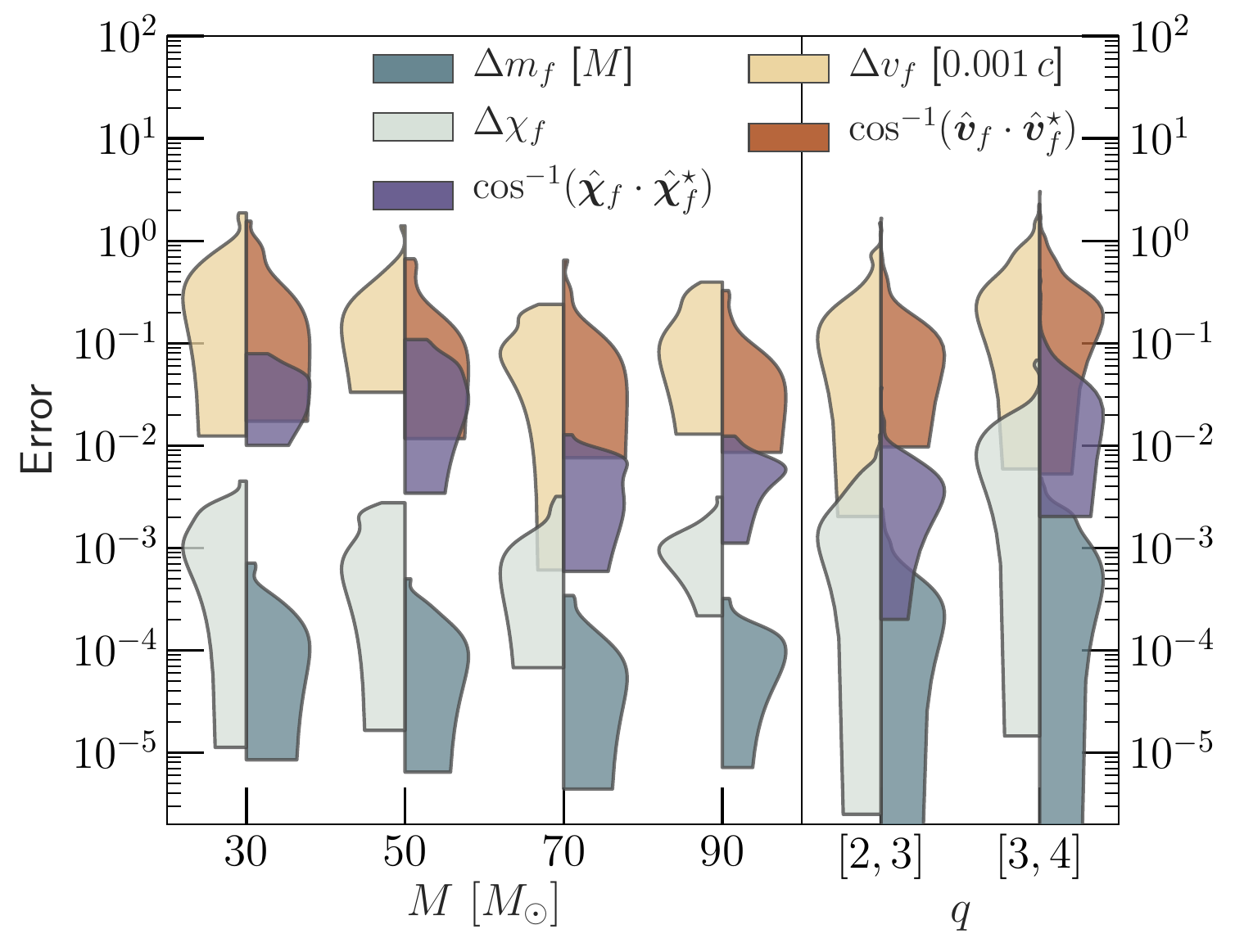}
\caption{Left panel: Errors for \PrecessingModelName in predicting remnant
properties when the spins are specified at an orbital frequency of
$f_{0}\!=\!10 \,{\rm Hz}$, for different total masses.
Right panel: Errors for
\PrecessingModelName when extrapolating to higher mass ratios, with the spins
specified at $t\!=\!-100M$.  The labels on the horizontal axis indicate the
range of mass ratios being tested. Note that the distributions in these plots
are normalized to have a fixed height, not fixed area.\vspace{-1em}
}
\label{fig:pn_and_extrap_errors}
\end{figure}

\prlsec{Regime of validity}
The errors in Fig.~\ref{fig:precessing_fits} are obtained by evaluating fits
using input spins specified at $t\!=\!-100M$, i.e., where the GPR interpolation
is performed.  The input spins can also be specified at earlier times; this
case is handled by two additional layers of time evolution.  Given the spins at
an initial orbital frequency $f_0$, we first evolve the spins using a
post-Newtonian (PN) approximant --- 3.5PN SpinTaylorT4
\cite{2003PhRvD..67j4025B,2007PhRvD..76l4038B,2015PhRvD..92j4028O}--- until the
orbital frequency reaches a value of $0.018\,\text{rad}/M$. At this point, we
are in the range of validity of the (more accurate) NRSur7dq2
approximant~\cite{2017PhRvD..96b4058B}, which we use to evolve the spins until
$t\!=\!-100 M$. Thus, spins can be specified at any given orbital frequency
and are evolved consistently before estimating the final BH properties.  This
is a crucial improvement over previous results, which, being calibrated solely
to non-precessing systems, suffer from ambiguities regarding the
separation/frequency at which spins are defined.

The left panel of Fig.~\ref{fig:pn_and_extrap_errors} shows the errors when the spins
are specified at an orbital frequency $f_0\!=\!10\, {\rm Hz}$. These errors are
computed by comparing against 20 long NR simulations~\cite{Catalog2018} with
mass ratios $q\leq2$ and generically oriented spins with magnitudes $\chi_1,
\chi_2 \leq 0.5$. None of these simulations were used to train the fits.  Longer PN evolutions are needed at lower total masses, and
the errors are therefore larger. These errors will decrease with an
improved spin evolution procedure. Note, however, that our predictions are still
more accurate (and, crucially, unambiguous) than those of existing fitting
formulae (cf. Fig.~\ref{fig:precessing_fits}).

Finally, the right panel of Fig.~\ref{fig:pn_and_extrap_errors} shows the the
performance of \PrecessingModelName when extrapolating to more extreme mass
ratios. We compare against 175 (225) NR simulations~\cite{Inprep-Varma:2019}
with $2\!\leq \!q \!\leq \!3$ ($3\!\leq\! q \!\leq \!4$), and generically
oriented spins with magnitudes $\chi_1, \chi_2 \leq 0.8$ specified at
$t\!=\!-100M$. The error distribution broadens, but our fits still provide a
reasonable estimate of the final remnant properties even far out of the training
parameter space. Detailed results on extrapolation accuracy are provided in
the supplemental materials~\cite{surfsupplement}.

\prlsec{Conclusion}
\label{sec:conclusion}
We have presented two highly accurate surrogate models for the remnant properties of
BH binaries. \PrecessingModelName (\AlignedModelName) is trained against 890
(104) NR simulations with mass ratios $q\leq2$ ($q\leq8$) and precessing
(aligned) spins with magnitude $\chi_{1},\chi_{2}\leq0.8$.  Both models
use GPR to provide fits for the remnant mass, spin, and kick velocity (both
magnitudes and directions). Our findings are implemented in a public Python
module named \emph{surfinBH} (details are provided in the supplemental
materials~\cite{surfsupplement}).

For aligned spins, errors in \AlignedModelName are comparable to existing
fitting formulae for the final mass and kick magnitude, while the spin is
predicted about 5 times more accurately.  For precessing systems, errors in
\PrecessingModelName  for final mass, spin magnitude, and kick magnitude are
lower than all existing models by at least an order of magnitude.  Crucially,
our fits are free from ambiguities regarding the time/frequency at which
precessing quantities are specified.  This is a point of major improvement over
previous models, which all fail to fully capture precession effects.

Is this increased accuracy necessary? For current events like GW150914, the estimated error in the
remnant properties are $\Delta m_f \roughly 0.1 M$ and $\Delta\chi_f \roughly
0.1$ \cite{2016PhRvX...6d1015A}. These measurements are currently dominated by statistical errors, as the
systematics introduced by existing fits used in the analysis are  $\Delta m_f
\roughly  5 \into 10^{-3}M$ and $\Delta\chi_f \roughly 2 \into 10^{-2}$ (see
$95^{\rm th}$ percentile values in Fig.~\ref{fig:precessing_fits}).  Because
statistical errors scale approximately linearly with the detector sensitivity
\cite{2008PhRvD..77d2001V}, we estimate that systematic errors in current models for
$\chi_{f}$ will start dominating over statistical uncertainties at
signal-to-noise ratios which are $\roughly 5$ times larger than that of
GW150914. This will happen sooner rather than later, with current interferometers expected to
reach their design sensitivity in a few years \cite{2018LRR....21....3A}, and future instruments
already being scheduled \cite{2017arXiv170200786A} or planned \cite{2010CQGra..27s4002P,2017CQGra..34d4001A}. Our fits, being
an order of magnitude more accurate (see Fig.~\ref{fig:precessing_fits}),
introduce systematic errors which are expected to be relevant only at SNRs
$\roughly 50$ times larger than that of GW150914. As shown above, errors are
largely dominated by the underlying NR resolution, not by our fitting
procedure. The inclusion of self-force evolutions
alongside NR  in the training dataset might also be exploited to improve
extrapolation performance at $q\gg 1$; we leave this to future work.

Moreover, the GPR methods employed here naturally provide error estimates along
with the fitted values (some results are provided in the supplemental
material~\cite{surfsupplement}). This constitutes a further key application of our results: when
performing, e.g., consistency tests of GR
\cite{2018CQGra..35a4002G,2016PhRvL.116v1101A}, systematic uncertainties
introduced by remnant fits can be naturally incorporated into the statistical
analysis and marginalized over (cf.  Ref.~\cite{2017PhRvD..96j2001C} for a
similar application of GPR and Refs.~\cite{2014PhRvL.113y1101M,2016PhRvD..93f4001M,2017PhRvD..96l3011D,2018PhRvD..97b4031H,2018arXiv180608365T} for other applications to GW science).

As GW astrophysics turns into a mature field, increasingly accurate tools such
as those presented here will become crucial to uncover more hidden secrets in
this new field of science.

\prlsec{Acknowledgments}
We thank Jonathan Blackman, Stephen Taylor, David Keitel, Anuradha Gupta, and
Serguei Ossokine for useful discussions. We made use of the public LIGO
Algorithm Library \cite{LAL} in the evaluation of existing fitting formulae and
to perform PN evolutions. We thank Nathan Johnson-McDaniel for useful
discussions, comments on the manuscript, and for sharing his code to evaluate
the HLZ kick fits.  V.V.~and F.H.~are supported by the Sherman Fairchild
Foundation and NSF grants PHY--1404569, PHY--170212, and PHY--1708213 at
Caltech.  D.G.~is supported by NASA through Einstein Postdoctoral Fellowship
Grant No.~PF6--170152 awarded by the Chandra X-ray Center, which is operated by
the Smithsonian Astrophysical Observatory for NASA under Contract NAS8-03060.
L.C.S.~acknowledges support from NSF grant PHY--1404569 and the Brinson
Foundation.  H.Z.~acknowledges support from the Caltech SURF Program and NSF
Grant No. PHY--1404569.  Computations were performed on NSF/NCSA Blue Waters
under allocation NSF PRAC--1713694 and on the Wheeler cluster at Caltech, which
is supported by the Sherman Fairchild Foundation and by Caltech.

%% file: supplement-content.tex
\setcounter{equation}{0}
\setcounter{figure}{0}
\setcounter{table}{0}
\renewcommand{\theequation}{S\arabic{equation}}
\renewcommand{\thefigure}{S\arabic{figure}}
\renewcommand{\bibnumfmt}[1]{[S#1]}
\renewcommand{\citenumfont}[1]{S#1}

\prlsec{Gaussian process regression}
\label{sec:gpr_fits}
We construct fits in this work using Gaussian process regression (GPR)
\cite{supp_2006gpml.book.....R, supp_2003itil.book.....M} as implemented in {\it
scikit-learn}~\cite{supp_2012arXiv1201.0490P}.

The starting point is a training set of $n$ observations, $\ts = \left\{(x^i,
f(x^i)) | i = 1,\dots,n\right\}$, where $x^i$ denotes an input vector of
dimension $D$ and $f(x^i)$ is the corresponding output. In our case, $x$
is mass ratio and spins of the merging binary, and $f(x)$ is
the remnant property we are fitting. Our goal is to use $\ts$ to make predictions for the
underlying $f(x)$ at any point $x_*$ that is not in $\ts$.

A Gaussian process (GP) can be thought of as a probability distribution of
functions. More formally, a GP is a collection of random variables, any finite
number of which have a joint Gaussian distribution~\cite{supp_2006gpml.book.....R}.  A GP is
completely specified by its mean function $m(x)$ and covariance function
$k(x,x^\prime)$, i.e.  $f(x) \sim \mathcal{GP}(m(x), k(x,x^\prime))$.  Consider
a prediction set of $n_*$ test inputs and their corresponding outputs (which
are unknown): $\ps = \left\{(x^i_*, f(x^i_*)) | i = 1,\dots,n_*\right\}$.  By
the definition of a GP, outputs of $\ts$ and $\ps$ (respectively ${\bf f}\!=\!\{f(x^i)\}$,
${\bf f_*}\!=\!\{f(x^i_*)\}$) are related by a joint
Gaussian distribution
\begin{gather}
\begin{bmatrix} {\bf{f}} \\ {\bf{f_*}} \end{bmatrix} =
    \mathcal{N}\left({\bf 0},
        \begin{bmatrix}
            K_{x x} & K_{x x_*} \\
            K_{x_* x} & K_{x_* x_*}
        \end{bmatrix}
    \right),
\label{Eq:GPR_prior}
\end{gather}
where $K_{x x_*}$ denotes the $n \into n_*$ matrix of the covariance
$k(x,x_*)$ evaluated at all pairs of training and prediction points, and
similarly for the other $K$ matrices.

Eq.~(\ref{Eq:GPR_prior}) provides the Bayesian prior distribution for ${\bf
f_*}$. The posterior distribution is obtained by restricting this joint prior
to contain only those functions which agree with the observed data points,
i.e.~\cite{supp_2006gpml.book.....R}
\begin{align}
p({\bf f_*}|\ts) = \mathcal{N} \Bigg( \! K_{x_* x} K_{x x}^{-1}
    {\bf f}, \, K_{x_* x_*} \!-\! K_{x_* x} K_{x x}^{-1} K_{x x_*} \! \Bigg).
\label{Eq:GPR_posterior}
\end{align}
The mean of this posterior provides an estimator for $f(x)$ at $x_*$, while its
width is the prediction error.

Finally, one needs to specify the covariance (or kernel) function
$k(x,x^\prime)$. In this \emph{Letter} we implement the following kernel
\begin{gather}
k(x, x^\prime) =\sigma_{k}^2 \exp{\left[ -\frac{1}{2}
\sum_{j=1}^{D}
\left(\frac{x^j - x^{\prime j}}{\sigma_j}\right)^2
\right]} + \sigma_n^2 \, \delta_{x,x'}\,,
\label{Eq:GPR_kernel}
\end{gather}
where $\delta_{x,x^{\prime}}$ is the Kronecker delta. In words, we use a
product between a squared exponential kernel and a constant kernel, to which we
add a white kernel term to account for additional noise in the training data
\cite{supp_2006gpml.book.....R,supp_2012arXiv1201.0490P}.

GPR fit construction involves determining the $D\!+\!2$ hyperparameters
($\sigma_k$, $\sigma_n$ and $\sigma_j$) which maximize the marginal likelihood
of the training data under the GP prior~\cite{supp_2006gpml.book.....R}. Local maxima are
avoided by repeating the optimization with 10 different initial guesses,
obtained by sampling uniformly in log in the hyperparameter space described
below.

Before constructing the GPR fit, we pre-process the training data as follows.
We first subtract a linear fit and the mean of the resulting values. Datapoints
are then normalized by dividing by the standard deviation of the resulting
values. The inverse of these transformations is applied at the time of the fit
evaluation.

For each dimension of $x$, we define $\Delta x^j$ to be the range of the values
of $x^j$ in $\ts$ and consider $\sigma_j \in [0.01 \times \Delta x^j, 10\times
\Delta x^j]$. Larger length scales are unlikely to be relevant and smaller
length scales are unlikely to be resolvable. The remaining hyperparameters are
sampled in $\sigma_k^2\in [10^{-2}, 10^2]$ and  $\sigma_n^2 \in [10^{-7},
10^{-2}]$. These choices are meant to be conservative and are based on prior
exploration of the typical magnitude and noise level in our pre-processed
training data.

\prlsec{Input parameter space}
Fits for \AlignedModelName are parameterized using $x = [\log(q), \hat{\chi},
\chi_a]$, where $\hat{\chi}$ is the spin parameter entering the GW phase at
leading order \cite{supp_2016PhRvD..93d4007K, supp_2011PhRvD..84h4037A, supp_1994PhRvD..49.2658C,
supp_1995PhRvD..52..848P} in the PN expansion,
\begin{gather}
\chieff = \frac{q~\chi_{1z} + \chi_{2z}}{1+q}\,, \qquad \eta = \frac{q}{(1+q)^2}\,,\\
\hat{\chi} = \frac{\chieff - 38\eta(\chi_{1z} + \chi_{2z})/113}
    {1-{76\eta}/{113}}\,, 
\end{gather}
and $\chi_a$ is the ``anti-symmetric spin'',
\begin{equation}
    \chi_a = \tfrac{1}{2}(\chi_{1z} - \chi_{2z})\,.
\label{eq:anti_symm_spin}
\end{equation}
For \PrecessingModelName we use $x=[\log(q), \chi_{1x}, \chi_{1y}, \hat{\chi},
\chi_{2x}, \chi_{2y}, \chi_{a}]$. Subscripts $x$, $y$ and $z$ refer to
components specified in the coorbital frame at $t\!=\!-100M$.  We empirically
found these parameterizations to perform more accurately than the more
intuitive choice $x=[q,\chi_{1x}, \chi_{1y}, \chi_{1z}, \chi_{2x}, \chi_{2y},
\chi_{2z}]$.

In the main text we describe how we evolve spins given at earlier times to
$t\!=\!-100M$, using PN and NRSur7qd2.  Is it worth noting that the NR spins
used to train \mbox{NRSur7qd2} had some additional smoothening filters applied
to them (see Eq.~6 in \cite{supp_2017PhRvD..96b4058B}). This introduces
additional systematics when evolving spins from times $t\!<\!-100M$. We
verified that the resulting errors on our fits are subdominant.

\begin{figure}[t]
\includegraphics[width=\columnwidth]{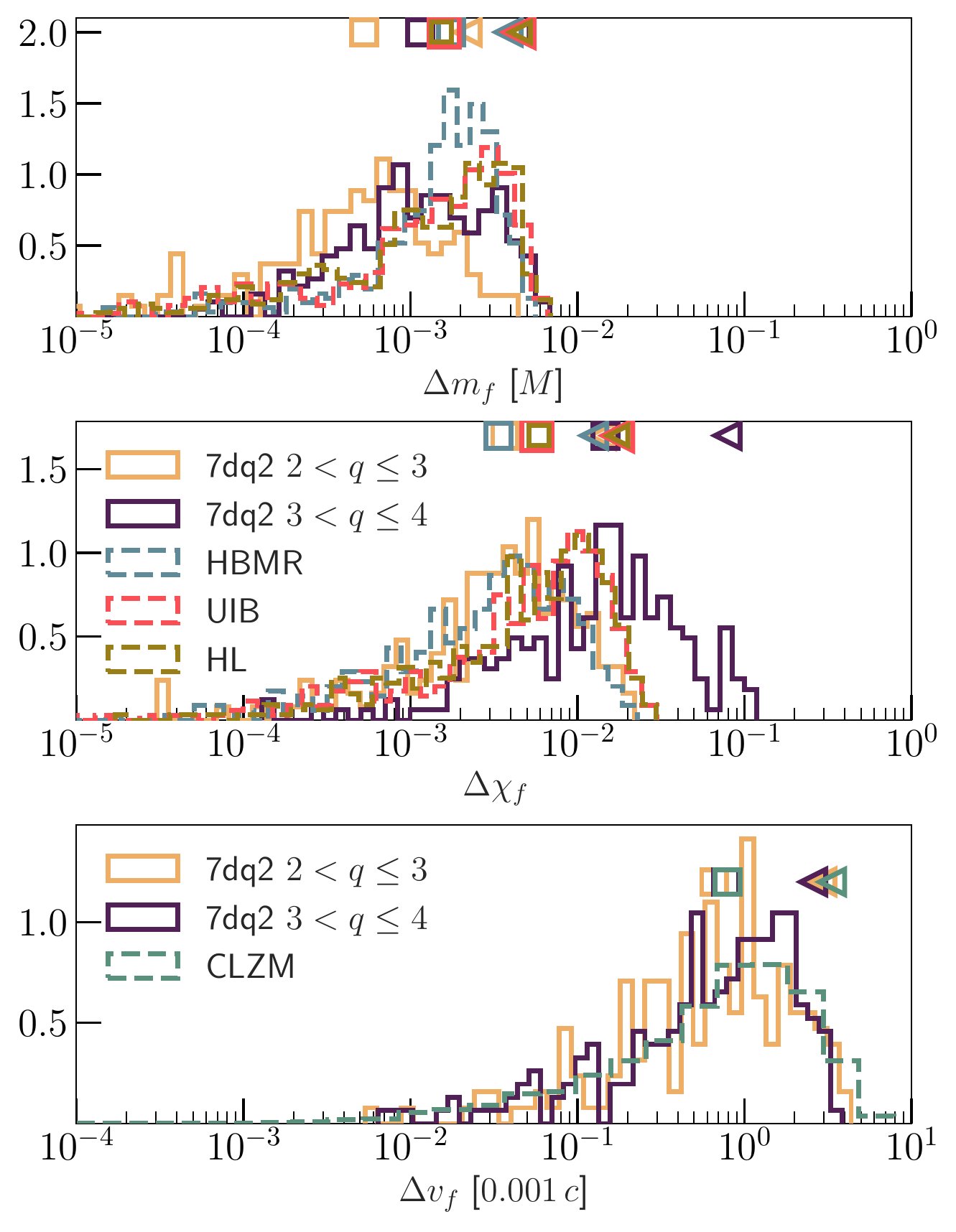}
\caption{Errors in \PrecessingModelName when extrapolating to higher mass
ratios, and the spins are specified at an orbital frequency
$f_0\!=\!10\, {\rm Hz}$, for a total mass $M=70 M_{\odot}$.
\vspace{-1em}
}
\label{fig:extrap_q4_and_NRSur_errs}
\end{figure}
\prlsec{Extrapolation erorrs}
\label{sec:app_extr_outs_param}
\begin{figure*}[p]
\includegraphics[scale=0.65]{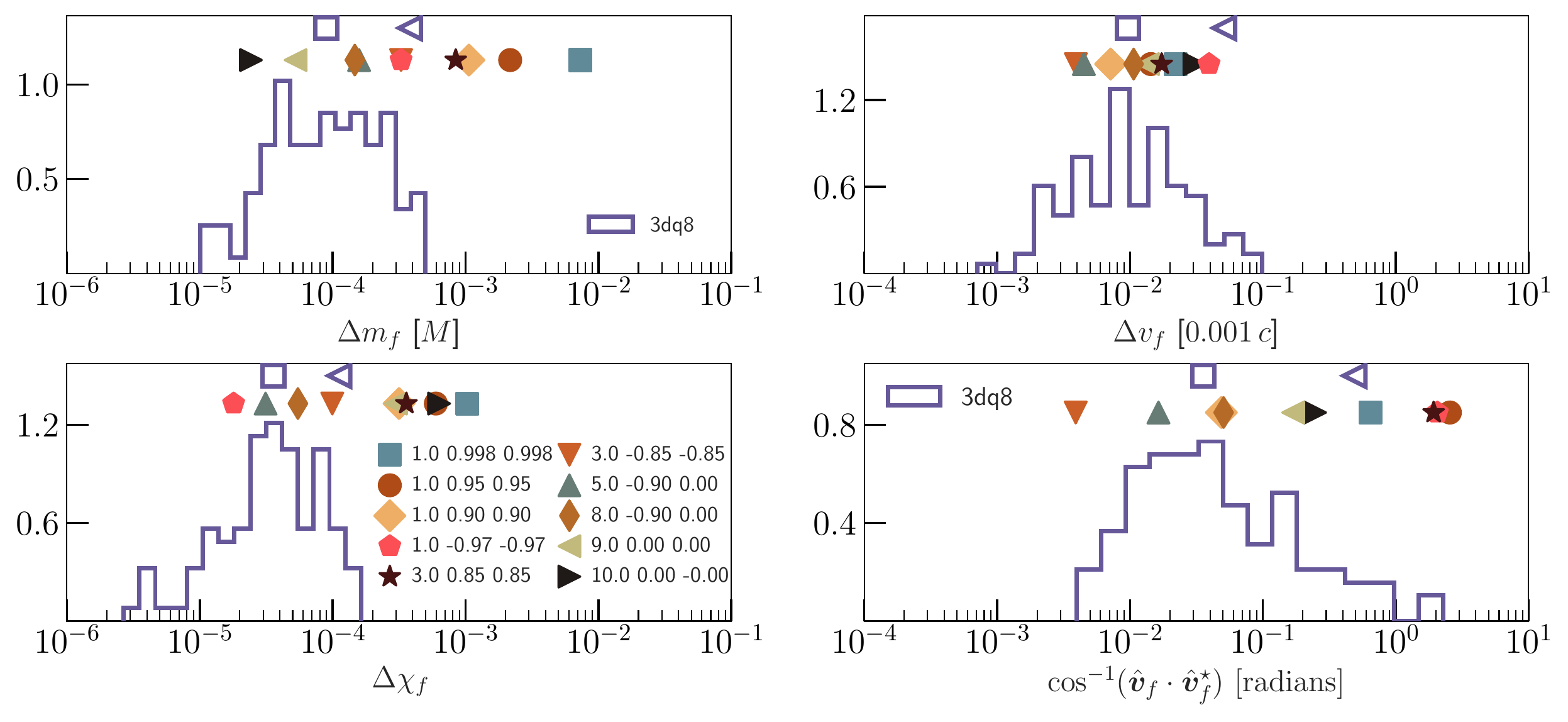}
\caption{Errors in predicting the remnant mass, spin, kick magnitude and kick
direction for nonprecessing BBH when  \AlignedModelName is extrapolated outside
of the training region (i.e.  $q>8$ and  $\chi_{1},\chi_{2}>0.8$). Each solid
symbol marks the error of the extrapolated model against a single nonprecessing
NR simulation. The legend in the bottom-left panel displays the mass ratio and
spin components of the two BHs along the orbital angular momentum direction.
Histograms of errors  within the training region (from
Fig.~2)
 are reproduced here for comparison.  The
hollow square (triangle) markers indicate the median ($95^{\rm th}$ percentile)
values for those errors.
}
\label{fig:extrap_aligned_spin_fits}
\vspace{1cm}
\includegraphics[scale=0.65]{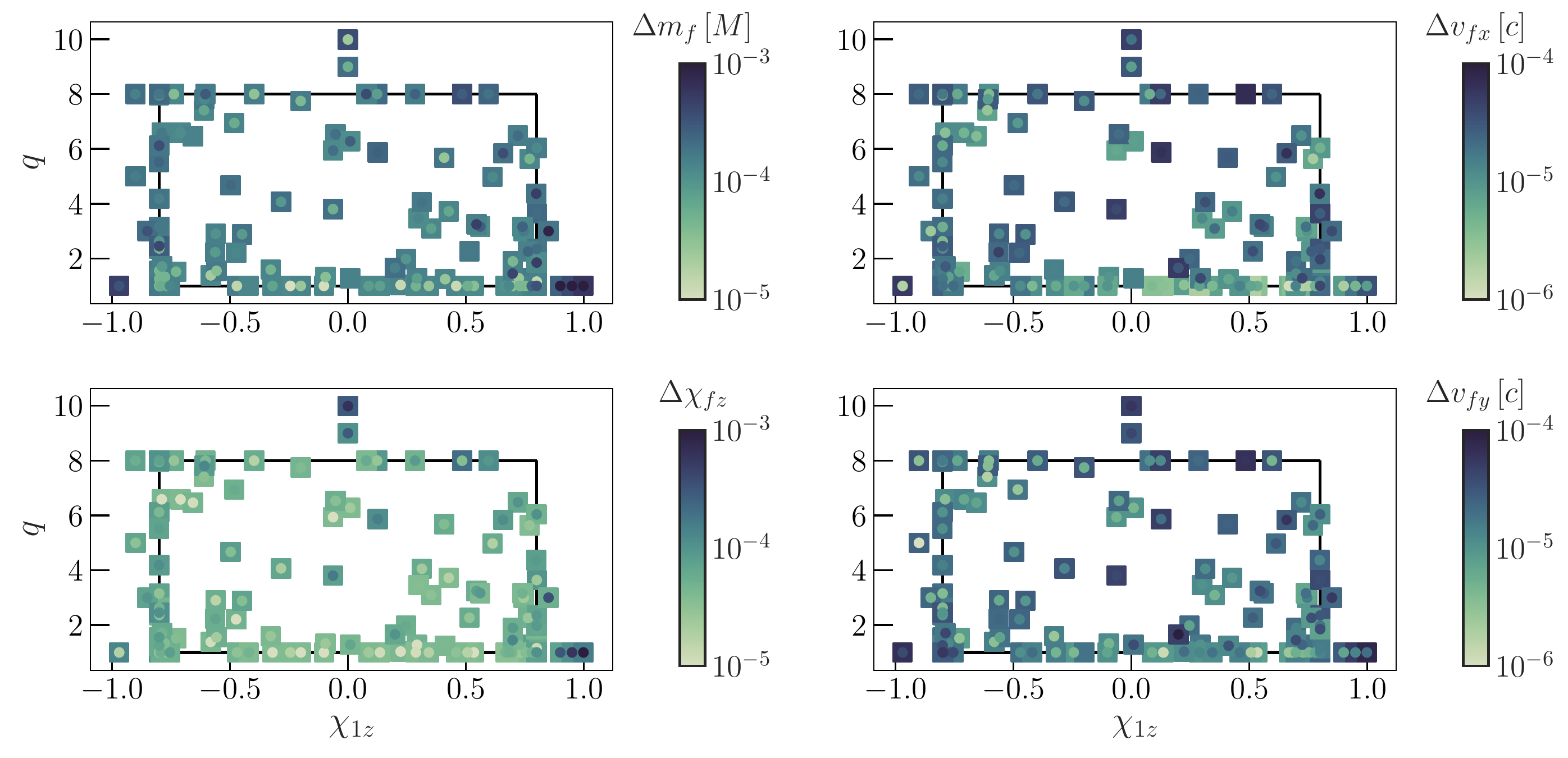}
\caption{Prediction errors for remnant mass, spin and kick for the model
\AlignedModelName against NR simulations. Two error estimates, as reported on
the color scale, are compared: out-of-sample errors marked with circles, and
$1\sigma$ GPR errors marked with squares.  We include cases where
\AlignedModelName needs to be extrapolated to higher mass ratios and/or spin
magnitudes. The bounds of the training parameter space are indicated as a black
rectangle.}
\label{fig:gpr_err_est_aligned_spin}
\end{figure*}
The right panel of Fig.~4 shows the errors in remnant quantities when
extrapolating \PrecessingModelName to mass ratios beyond its training range
($q\leq2$). These errors are computed using the spins at $t\!=\!-100M$. If the
spins are given at earlier times, we expect larger extrapolation errors as this
also involves extrapolation of the NRSur7dq2 waveform model (which was also
trained in the $q\leq2$ space). Figure \ref{fig:extrap_q4_and_NRSur_errs}
shows the extrapolation errors when the spins are specified at at orbital
frequency $f_0\!=\!10\, {\rm Hz}$ for a total mass $M=70 M_{\odot}$, computed by comparing against the same NR simulations as in Fig.~4.
Errors are comparable to or lower than those of existing fits for $q\leq3$.
For $3<q\leq4$, our errors for the remnant spin magnitude can become larger, but
the remnant mass and kick magnitude remains as accurate as in other
fits.

Figure \ref{fig:extrap_aligned_spin_fits} shows errors in \AlignedModelName
when extrapolated beyond its training space to higher mass ratios and/or spin
magnitudes (this figure complements the results shown in Fig.~4 of the
main text
for \PrecessingModelName). Here we used some of the simulations
of~\cite{supp_2013PhRvL.111x1104M, supp_2015PhRvD..92j2001K,
supp_2015PhRvL.115l1102B, supp_2016CQGra..33p5001C, supp_Catalog2018} with
$q>8$ and/or $\chi_1,\chi_2>0.8$.  Accuracy in the remnant mass degrades
noticeably only at high ($\sim 0.9$) co-aligned spins. Errors in final spin
become larger at both high spins and extreme mass ratios. For counter-aligned
spins, our errors are always comparable to those found within the training
region. Errors in kick magnitude and direction appear to be insensitive to
extrapolation.

\prlsec{GPR error prediction}
\begin{figure*}[thb]
\vspace*{-1cm}
\includegraphics[scale=0.7]{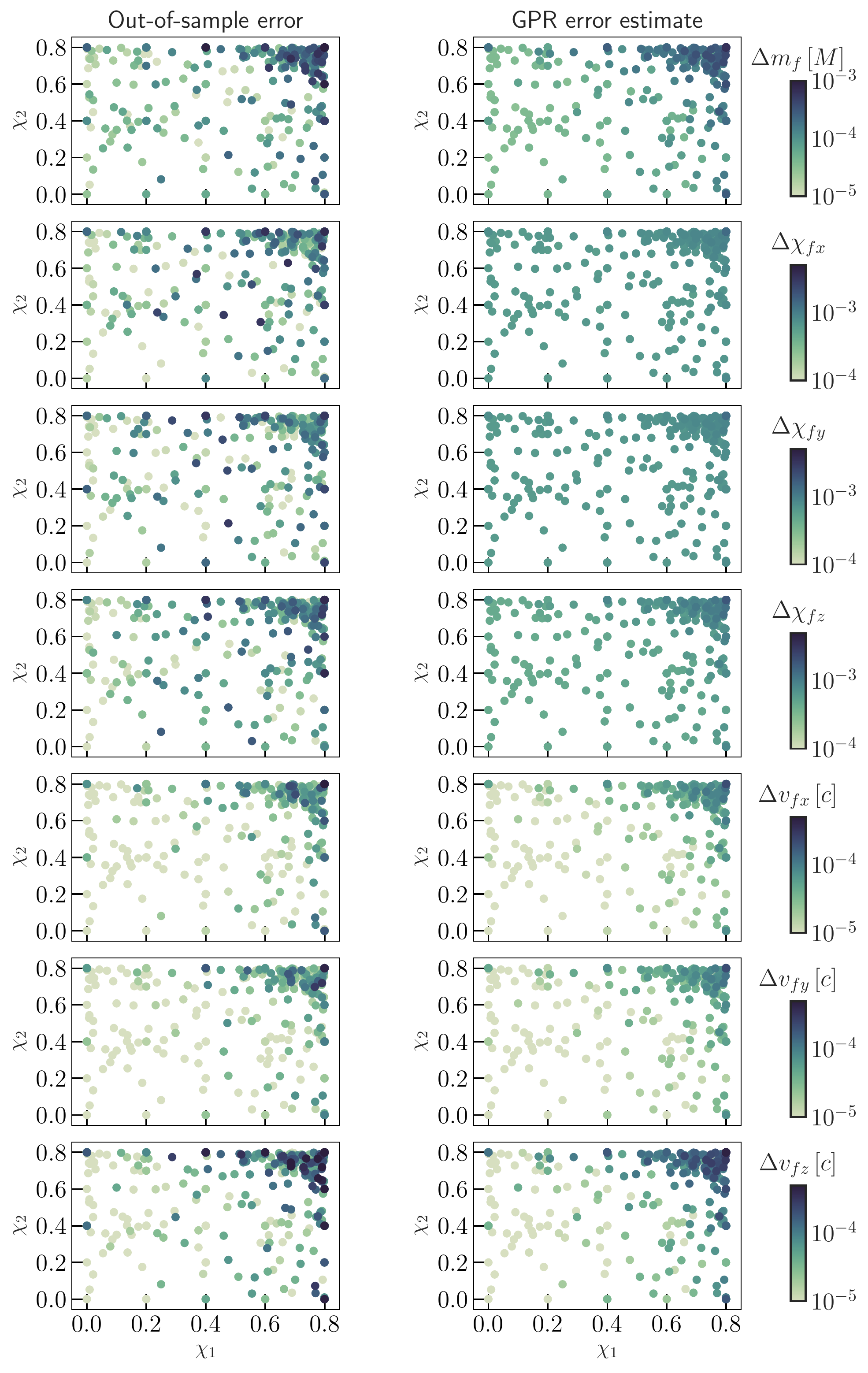}
\vspace*{-0.2cm}
\caption{Comparison between out-of-sample (left) and $1\sigma$ GPR (right)
errors for \PrecessingModelName. The axes show the magnitudes of the component
spins, and the color scale indicates the parameter error being plotted.
}
\label{fig:gpr_err_est_precessing}
\end{figure*}
As stressed above and in the main body of our \emph{Letter}, GPR naturally
associates errors to the estimated quantities. In this Section we test the
efficacy of this prediction by comparing the GPR errors against the
out-of-sample errors. The GPR errors shown here are evaluated using the same
cross-validation data sets used to generate the out-of-sample errors.
Therefore, both error estimates are evaluated at points in parameter space
where models were not trained.

Error comparisons for \AlignedModelName and \PrecessingModelName are reported
in Figs.~\ref{fig:gpr_err_est_aligned_spin} and
\ref{fig:gpr_err_est_precessing}, respectively.  While GPR predictions miss
some of the features captured by the ``k-fold'' cross validations, overall it
provides faithful estimates of the fit errors.

\prlsec{Public python implementation}
Our fits are made publicly available through the easy-to-use Python package,
\PackageName~\cite{supp_surfinBH}. Our code is compatible with both \texttt{Python~2} and \texttt{Python~3}. The latest release can be installed from the Python
Package Index using
\begin{verbatim}
    pip install surfinBH
\end{verbatim}

Python packages {\it numpy}~\cite{supp_Walt}, {\it scipy}~\cite{supp_Jones:2001aa}, {\it h5py}~\cite{supp_h5py},
{\it scikit-learn}~\cite{supp_2012arXiv1201.0490P}, {\it lalsuite}~\cite{supp_LAL}, and
{\it NRSur7dq2}~\cite{supp_2017PhRvD..96b4058B} are specified as dependencies and are
automatically installed if missing.  \PackageName is hosted on GitHub at
\href{https://github.com/vijayvarma392/surfinBH}{github.com/vijayvarma392/surfinBH},
from which development versions can be installed. Continuous integration is
provided by  \textit{Travis}~\cite{supp_travis-ci}

The \PackageName module can be imported in Python using
\begin{verbatim}
    import surfinBH
\end{verbatim}
Documentation is provided for each submodule of surfinBH and can be accessed
via Python's \texttt{help()} function. The fit class has to be initialized
using, e.g.
\begin{verbatim}
    fit = surfinBH.LoadFits("surfinBH7dq2")
\end{verbatim}
Given mass ratio and component spins, the fits and $1\sigma$ GPR error
estimates of the remnant mass, spin vector and kick vector can be evaluated as
follows:
\begin{verbatim}
    q = 1.2
    chiA = [0.5, 0.05, 0.3]
    chiB = [-0.5, -0.05, 0.1]
    mf, mf_err = fit.mf(q, chiA, chiB)
    chif, chif_err = fit.chif(q, chiA, chiB)
    vf, vf_err = fit.vf(q, chiA, chiB)
\end{verbatim}
Both the input spins as well as the remnant spin and kick vectors are assumed
to be specified in the coorbital frame at $t\!=\!-100M$.  Performance of
\PackageName was tested on a 3.1\,GHz {Intel Core i5} processor by averaging
over $10^3$ evaluations at randomly chosen points in parameter space. For
\PrecessingModelName, evaluation cost of final mass (spin) [kick] is 2.5\,ms
(7\,ms) [7\,ms]. For \AlignedModelName, evaluation cost of final mass (spin)
[kick] is 0.4\,ms (0.4\,ms) [0.6\,ms].

We also allow specifying an orbital frequency (in units of $\text{rad}/M$),
e.g.:
\begin{verbatim}
    omega0 = 5e-3
    mf, mf_err = fit.mf(q, chiA, chiB,
                        omega0 = omega0)
    chif, chif_err = fit.chif(q, chiA, chiB,
                              omega0 = omega0)
    vf, vf_err = fit.vf(q, chiA, chiB,
                        omega0 = omega0)
\end{verbatim}
In this case, the component spins, as well as the final remnant spin/kick
vectors are specified in the coorbital frame at this orbital frequency.  The
evaluation costs are larger when specifying an initial orbital frequency since
this involves two additional stages of spin evolution. Execution times depend
on the initial frequency, the specific PN approximant used and the time step
size in the integration routine. For instance, with \texttt{omega0 = 5e-3},
SpinTaylorT4, and a step size of $0.1 M$ the evaluation cost is $\roughly 0.5$s
for each of the remnant quantities.

Additional resources are provided in the package installation
page~\cite{supp_surfinBH}. This includes example \texttt{jupyter} notebooks for both
models presented in this \emph{Letter}.

%% file: finalfits.bbl
\begin{thebibliography}{87}%
\makeatletter
\providecommand \@ifxundefined [1]{%
 \@ifx{#1\undefined}
}%
\providecommand \@ifnum [1]{%
 \ifnum #1\expandafter \@firstoftwo
 \else \expandafter \@secondoftwo
 \fi
}%
\providecommand \@ifx [1]{%
 \ifx #1\expandafter \@firstoftwo
 \else \expandafter \@secondoftwo
 \fi
}%
\providecommand \natexlab [1]{#1}%
\providecommand \enquote  [1]{``#1''}%
\providecommand \bibnamefont  [1]{#1}%
\providecommand \bibfnamefont [1]{#1}%
\providecommand \citenamefont [1]{#1}%
\providecommand \href@noop [0]{\@secondoftwo}%
\providecommand \href [0]{\begingroup \@sanitize@url \@href}%
\providecommand \@href[1]{\@@startlink{#1}\@@href}%
\providecommand \@@href[1]{\endgroup#1\@@endlink}%
\providecommand \@sanitize@url [0]{\catcode `\\12\catcode `\$12\catcode
  `\&12\catcode `\#12\catcode `\^12\catcode `\_12\catcode `\%12\relax}%
\providecommand \@@startlink[1]{}%
\providecommand \@@endlink[0]{}%
\providecommand \url  [0]{\begingroup\@sanitize@url \@url }%
\providecommand \@url [1]{\endgroup\@href {#1}{\urlprefix }}%
\providecommand \urlprefix  [0]{URL }%
\providecommand \Eprint [0]{\href }%
\providecommand \doibase [0]{http://dx.doi.org/}%
\providecommand \selectlanguage [0]{\@gobble}%
\providecommand \bibinfo  [0]{\@secondoftwo}%
\providecommand \bibfield  [0]{\@secondoftwo}%
\providecommand \translation [1]{[#1]}%
\providecommand \BibitemOpen [0]{}%
\providecommand \bibitemStop [0]{}%
\providecommand \bibitemNoStop [0]{.\EOS\space}%
\providecommand \EOS [0]{\spacefactor3000\relax}%
\providecommand \BibitemShut  [1]{\csname bibitem#1\endcsname}%
\let\auto@bib@innerbib\@empty
\bibitem [{\citenamefont {{Israel}}(1968)}]{1968CMaPh...8..245I}%
  \BibitemOpen
  \bibfield  {author} {\bibinfo {author} {\bibfnamefont {W.}~\bibnamefont
  {{Israel}}},\ }\href {\doibase 10.1007/BF01645859} {\bibfield  {journal}
  {\bibinfo  {journal} {Communications in Mathematical Physics}\ }\textbf
  {\bibinfo {volume} {8}},\ \bibinfo {pages} {245} (\bibinfo {year}
  {1968})}\BibitemShut {NoStop}%
\bibitem [{\citenamefont {{Carter}}(1971)}]{1971PhRvL..26..331C}%
  \BibitemOpen
  \bibfield  {author} {\bibinfo {author} {\bibfnamefont {B.}~\bibnamefont
  {{Carter}}},\ }\href {\doibase 10.1103/PhysRevLett.26.331} {\bibfield
  {journal} {\bibinfo  {journal} {\prl}\ }\textbf {\bibinfo {volume} {26}},\
  \bibinfo {pages} {331} (\bibinfo {year} {1971})}\BibitemShut {NoStop}%
\bibitem [{\citenamefont {{Heusler}}(1996)}]{1996bhut.book.....H}%
  \BibitemOpen
  \bibfield  {author} {\bibinfo {author} {\bibfnamefont {M.}~\bibnamefont
  {{Heusler}}},\ }\href@noop {} {\emph {\bibinfo {title} {{Black hole
  uniqueness theorems}, Cambridge University Press, 1996.}}}\ (\bibinfo {year}
  {1996})\BibitemShut {NoStop}%
\bibitem [{\citenamefont {{Hannam}}\ \emph {et~al.}(2014)\citenamefont
  {{Hannam}}, \citenamefont {{Schmidt}}, \citenamefont {{Boh{\'e}}},
  \citenamefont {{Haegel}}, \citenamefont {{Husa}}, \citenamefont {{Ohme}},
  \citenamefont {{Pratten}},\ and\ \citenamefont
  {{P{\"u}rrer}}}]{2014PhRvL.113o1101H}%
  \BibitemOpen
  \bibfield  {author} {\bibinfo {author} {\bibfnamefont {M.}~\bibnamefont
  {{Hannam}}}, \bibinfo {author} {\bibfnamefont {P.}~\bibnamefont {{Schmidt}}},
  \bibinfo {author} {\bibfnamefont {A.}~\bibnamefont {{Boh{\'e}}}}, \bibinfo
  {author} {\bibfnamefont {L.}~\bibnamefont {{Haegel}}}, \bibinfo {author}
  {\bibfnamefont {S.}~\bibnamefont {{Husa}}}, \bibinfo {author} {\bibfnamefont
  {F.}~\bibnamefont {{Ohme}}}, \bibinfo {author} {\bibfnamefont
  {G.}~\bibnamefont {{Pratten}}}, \ and\ \bibinfo {author} {\bibfnamefont
  {M.}~\bibnamefont {{P{\"u}rrer}}},\ }\href {\doibase
  10.1103/PhysRevLett.113.151101} {\bibfield  {journal} {\bibinfo  {journal}
  {\prl}\ }\textbf {\bibinfo {volume} {113}},\ \bibinfo {eid} {151101}
  (\bibinfo {year} {2014})},\ \Eprint {http://arxiv.org/abs/1308.3271}
  {arXiv:1308.3271 [gr-qc]} \BibitemShut {NoStop}%
\bibitem [{\citenamefont {{Khan}}\ \emph {et~al.}(2016)\citenamefont {{Khan}},
  \citenamefont {{Husa}}, \citenamefont {{Hannam}}, \citenamefont {{Ohme}},
  \citenamefont {{P{\"u}rrer}}, \citenamefont {{Forteza}},\ and\ \citenamefont
  {{Boh{\'e}}}}]{2016PhRvD..93d4007K}%
  \BibitemOpen
  \bibfield  {author} {\bibinfo {author} {\bibfnamefont {S.}~\bibnamefont
  {{Khan}}}, \bibinfo {author} {\bibfnamefont {S.}~\bibnamefont {{Husa}}},
  \bibinfo {author} {\bibfnamefont {M.}~\bibnamefont {{Hannam}}}, \bibinfo
  {author} {\bibfnamefont {F.}~\bibnamefont {{Ohme}}}, \bibinfo {author}
  {\bibfnamefont {M.}~\bibnamefont {{P{\"u}rrer}}}, \bibinfo {author}
  {\bibfnamefont {X.~J.}\ \bibnamefont {{Forteza}}}, \ and\ \bibinfo {author}
  {\bibfnamefont {A.}~\bibnamefont {{Boh{\'e}}}},\ }\href {\doibase
  10.1103/PhysRevD.93.044007} {\bibfield  {journal} {\bibinfo  {journal}
  {\prd}\ }\textbf {\bibinfo {volume} {93}},\ \bibinfo {eid} {044007} (\bibinfo
  {year} {2016})},\ \Eprint {http://arxiv.org/abs/1508.07253} {arXiv:1508.07253
  [gr-qc]} \BibitemShut {NoStop}%
\bibitem [{\citenamefont {{Husa}}\ \emph {et~al.}(2016)\citenamefont {{Husa}},
  \citenamefont {{Khan}}, \citenamefont {{Hannam}}, \citenamefont
  {{P{\"u}rrer}}, \citenamefont {{Ohme}}, \citenamefont {{Forteza}},\ and\
  \citenamefont {{Boh{\'e}}}}]{2016PhRvD..93d4006H}%
  \BibitemOpen
  \bibfield  {author} {\bibinfo {author} {\bibfnamefont {S.}~\bibnamefont
  {{Husa}}}, \bibinfo {author} {\bibfnamefont {S.}~\bibnamefont {{Khan}}},
  \bibinfo {author} {\bibfnamefont {M.}~\bibnamefont {{Hannam}}}, \bibinfo
  {author} {\bibfnamefont {M.}~\bibnamefont {{P{\"u}rrer}}}, \bibinfo {author}
  {\bibfnamefont {F.}~\bibnamefont {{Ohme}}}, \bibinfo {author} {\bibfnamefont
  {X.~J.}\ \bibnamefont {{Forteza}}}, \ and\ \bibinfo {author} {\bibfnamefont
  {A.}~\bibnamefont {{Boh{\'e}}}},\ }\href {\doibase
  10.1103/PhysRevD.93.044006} {\bibfield  {journal} {\bibinfo  {journal}
  {\prd}\ }\textbf {\bibinfo {volume} {93}},\ \bibinfo {eid} {044006} (\bibinfo
  {year} {2016})},\ \Eprint {http://arxiv.org/abs/1508.07250} {arXiv:1508.07250
  [gr-qc]} \BibitemShut {NoStop}%
\bibitem [{\citenamefont {{Buonanno}}\ \emph {et~al.}(2009)\citenamefont
  {{Buonanno}}, \citenamefont {{Pan}}, \citenamefont {{Pfeiffer}},
  \citenamefont {{Scheel}}, \citenamefont {{Buchman}},\ and\ \citenamefont
  {{Kidder}}}]{2009PhRvD..79l4028B}%
  \BibitemOpen
  \bibfield  {author} {\bibinfo {author} {\bibfnamefont {A.}~\bibnamefont
  {{Buonanno}}}, \bibinfo {author} {\bibfnamefont {Y.}~\bibnamefont {{Pan}}},
  \bibinfo {author} {\bibfnamefont {H.~P.}\ \bibnamefont {{Pfeiffer}}},
  \bibinfo {author} {\bibfnamefont {M.~A.}\ \bibnamefont {{Scheel}}}, \bibinfo
  {author} {\bibfnamefont {L.~T.}\ \bibnamefont {{Buchman}}}, \ and\ \bibinfo
  {author} {\bibfnamefont {L.~E.}\ \bibnamefont {{Kidder}}},\ }\href {\doibase
  10.1103/PhysRevD.79.124028} {\bibfield  {journal} {\bibinfo  {journal}
  {\prd}\ }\textbf {\bibinfo {volume} {79}},\ \bibinfo {eid} {124028} (\bibinfo
  {year} {2009})},\ \Eprint {http://arxiv.org/abs/0902.0790} {arXiv:0902.0790
  [gr-qc]} \BibitemShut {NoStop}%
\bibitem [{\citenamefont {{Boh{\'e}}}\ \emph {et~al.}(2017)\citenamefont
  {{Boh{\'e}}}, \citenamefont {{Shao}}, \citenamefont {{Taracchini}},
  \citenamefont {{Buonanno}}, \citenamefont {{Babak}}, \citenamefont {{Harry}},
  \citenamefont {{Hinder}}, \citenamefont {{Ossokine}}, \citenamefont
  {{P{\"u}rrer}}, \citenamefont {{Raymond}}, \citenamefont {{Chu}},
  \citenamefont {{Fong}}, \citenamefont {{Kumar}}, \citenamefont {{Pfeiffer}},
  \citenamefont {{Boyle}}, \citenamefont {{Hemberger}}, \citenamefont
  {{Kidder}}, \citenamefont {{Lovelace}}, \citenamefont {{Scheel}},\ and\
  \citenamefont {{Szil{\'a}gyi}}}]{2017PhRvD..95d4028B}%
  \BibitemOpen
  \bibfield  {author} {\bibinfo {author} {\bibfnamefont {A.}~\bibnamefont
  {{Boh{\'e}}}}, \bibinfo {author} {\bibfnamefont {L.}~\bibnamefont {{Shao}}},
  \bibinfo {author} {\bibfnamefont {A.}~\bibnamefont {{Taracchini}}}, \bibinfo
  {author} {\bibfnamefont {A.}~\bibnamefont {{Buonanno}}}, \bibinfo {author}
  {\bibfnamefont {S.}~\bibnamefont {{Babak}}}, \bibinfo {author} {\bibfnamefont
  {I.~W.}\ \bibnamefont {{Harry}}}, \bibinfo {author} {\bibfnamefont
  {I.}~\bibnamefont {{Hinder}}}, \bibinfo {author} {\bibfnamefont
  {S.}~\bibnamefont {{Ossokine}}}, \bibinfo {author} {\bibfnamefont
  {M.}~\bibnamefont {{P{\"u}rrer}}}, \bibinfo {author} {\bibfnamefont
  {V.}~\bibnamefont {{Raymond}}}, \bibinfo {author} {\bibfnamefont
  {T.}~\bibnamefont {{Chu}}}, \bibinfo {author} {\bibfnamefont
  {H.}~\bibnamefont {{Fong}}}, \bibinfo {author} {\bibfnamefont
  {P.}~\bibnamefont {{Kumar}}}, \bibinfo {author} {\bibfnamefont {H.~P.}\
  \bibnamefont {{Pfeiffer}}}, \bibinfo {author} {\bibfnamefont
  {M.}~\bibnamefont {{Boyle}}}, \bibinfo {author} {\bibfnamefont {D.~A.}\
  \bibnamefont {{Hemberger}}}, \bibinfo {author} {\bibfnamefont {L.~E.}\
  \bibnamefont {{Kidder}}}, \bibinfo {author} {\bibfnamefont {G.}~\bibnamefont
  {{Lovelace}}}, \bibinfo {author} {\bibfnamefont {M.~A.}\ \bibnamefont
  {{Scheel}}}, \ and\ \bibinfo {author} {\bibfnamefont {B.}~\bibnamefont
  {{Szil{\'a}gyi}}},\ }\href {\doibase 10.1103/PhysRevD.95.044028} {\bibfield
  {journal} {\bibinfo  {journal} {\prd}\ }\textbf {\bibinfo {volume} {95}},\
  \bibinfo {eid} {044028} (\bibinfo {year} {2017})},\ \Eprint
  {http://arxiv.org/abs/1611.03703} {arXiv:1611.03703 [gr-qc]} \BibitemShut
  {NoStop}%
\bibitem [{\citenamefont {{Babak}}\ \emph {et~al.}(2017)\citenamefont
  {{Babak}}, \citenamefont {{Taracchini}},\ and\ \citenamefont
  {{Buonanno}}}]{2017PhRvD..95b4010B}%
  \BibitemOpen
  \bibfield  {author} {\bibinfo {author} {\bibfnamefont {S.}~\bibnamefont
  {{Babak}}}, \bibinfo {author} {\bibfnamefont {A.}~\bibnamefont
  {{Taracchini}}}, \ and\ \bibinfo {author} {\bibfnamefont {A.}~\bibnamefont
  {{Buonanno}}},\ }\href {\doibase 10.1103/PhysRevD.95.024010} {\bibfield
  {journal} {\bibinfo  {journal} {\prd}\ }\textbf {\bibinfo {volume} {95}},\
  \bibinfo {eid} {024010} (\bibinfo {year} {2017})},\ \Eprint
  {http://arxiv.org/abs/1607.05661} {arXiv:1607.05661 [gr-qc]} \BibitemShut
  {NoStop}%
\bibitem [{\citenamefont {{Field}}\ \emph {et~al.}(2014)\citenamefont
  {{Field}}, \citenamefont {{Galley}}, \citenamefont {{Hesthaven}},
  \citenamefont {{Kaye}},\ and\ \citenamefont
  {{Tiglio}}}]{2014PhRvX...4c1006F}%
  \BibitemOpen
  \bibfield  {author} {\bibinfo {author} {\bibfnamefont {S.~E.}\ \bibnamefont
  {{Field}}}, \bibinfo {author} {\bibfnamefont {C.~R.}\ \bibnamefont
  {{Galley}}}, \bibinfo {author} {\bibfnamefont {J.~S.}\ \bibnamefont
  {{Hesthaven}}}, \bibinfo {author} {\bibfnamefont {J.}~\bibnamefont {{Kaye}}},
  \ and\ \bibinfo {author} {\bibfnamefont {M.}~\bibnamefont {{Tiglio}}},\
  }\href {\doibase 10.1103/PhysRevX.4.031006} {\bibfield  {journal} {\bibinfo
  {journal} {\prx}\ }\textbf {\bibinfo {volume} {4}},\ \bibinfo {eid} {031006}
  (\bibinfo {year} {2014})},\ \Eprint {http://arxiv.org/abs/1308.3565}
  {arXiv:1308.3565 [gr-qc]} \BibitemShut {NoStop}%
\bibitem [{\citenamefont {{Blackman}}\ \emph
  {et~al.}(2017{\natexlab{a}})\citenamefont {{Blackman}}, \citenamefont
  {{Field}}, \citenamefont {{Scheel}}, \citenamefont {{Galley}}, \citenamefont
  {{Hemberger}}, \citenamefont {{Schmidt}},\ and\ \citenamefont
  {{Smith}}}]{2017PhRvD..95j4023B}%
  \BibitemOpen
  \bibfield  {author} {\bibinfo {author} {\bibfnamefont {J.}~\bibnamefont
  {{Blackman}}}, \bibinfo {author} {\bibfnamefont {S.~E.}\ \bibnamefont
  {{Field}}}, \bibinfo {author} {\bibfnamefont {M.~A.}\ \bibnamefont
  {{Scheel}}}, \bibinfo {author} {\bibfnamefont {C.~R.}\ \bibnamefont
  {{Galley}}}, \bibinfo {author} {\bibfnamefont {D.~A.}\ \bibnamefont
  {{Hemberger}}}, \bibinfo {author} {\bibfnamefont {P.}~\bibnamefont
  {{Schmidt}}}, \ and\ \bibinfo {author} {\bibfnamefont {R.}~\bibnamefont
  {{Smith}}},\ }\href {\doibase 10.1103/PhysRevD.95.104023} {\bibfield
  {journal} {\bibinfo  {journal} {\prd}\ }\textbf {\bibinfo {volume} {95}},\
  \bibinfo {eid} {104023} (\bibinfo {year} {2017}{\natexlab{a}})},\ \Eprint
  {http://arxiv.org/abs/1701.00550} {arXiv:1701.00550 [gr-qc]} \BibitemShut
  {NoStop}%
\bibitem [{\citenamefont {{Blackman}}\ \emph
  {et~al.}(2017{\natexlab{b}})\citenamefont {{Blackman}}, \citenamefont
  {{Field}}, \citenamefont {{Scheel}}, \citenamefont {{Galley}}, \citenamefont
  {{Ott}}, \citenamefont {{Boyle}}, \citenamefont {{Kidder}}, \citenamefont
  {{Pfeiffer}},\ and\ \citenamefont {{Szil{\'a}gyi}}}]{2017PhRvD..96b4058B}%
  \BibitemOpen
  \bibfield  {author} {\bibinfo {author} {\bibfnamefont {J.}~\bibnamefont
  {{Blackman}}}, \bibinfo {author} {\bibfnamefont {S.~E.}\ \bibnamefont
  {{Field}}}, \bibinfo {author} {\bibfnamefont {M.~A.}\ \bibnamefont
  {{Scheel}}}, \bibinfo {author} {\bibfnamefont {C.~R.}\ \bibnamefont
  {{Galley}}}, \bibinfo {author} {\bibfnamefont {C.~D.}\ \bibnamefont {{Ott}}},
  \bibinfo {author} {\bibfnamefont {M.}~\bibnamefont {{Boyle}}}, \bibinfo
  {author} {\bibfnamefont {L.~E.}\ \bibnamefont {{Kidder}}}, \bibinfo {author}
  {\bibfnamefont {H.~P.}\ \bibnamefont {{Pfeiffer}}}, \ and\ \bibinfo {author}
  {\bibfnamefont {B.}~\bibnamefont {{Szil{\'a}gyi}}},\ }\href {\doibase
  10.1103/PhysRevD.96.024058} {\bibfield  {journal} {\bibinfo  {journal}
  {\prd}\ }\textbf {\bibinfo {volume} {96}},\ \bibinfo {eid} {024058} (\bibinfo
  {year} {2017}{\natexlab{b}})},\ \Eprint {http://arxiv.org/abs/1705.07089}
  {arXiv:1705.07089 [gr-qc]} \BibitemShut {NoStop}%
\bibitem [{\citenamefont {{Ghosh}}\ \emph {et~al.}(2018)\citenamefont
  {{Ghosh}}, \citenamefont {{Johnson-McDaniel}}, \citenamefont {{Ghosh}},
  \citenamefont {{Kant Mishra}}, \citenamefont {{Ajith}}, \citenamefont {{Del
  Pozzo}}, \citenamefont {{Berry}}, \citenamefont {{Nielsen}},\ and\
  \citenamefont {{London}}}]{2018CQGra..35a4002G}%
  \BibitemOpen
  \bibfield  {author} {\bibinfo {author} {\bibfnamefont {A.}~\bibnamefont
  {{Ghosh}}}, \bibinfo {author} {\bibfnamefont {N.~K.}\ \bibnamefont
  {{Johnson-McDaniel}}}, \bibinfo {author} {\bibfnamefont {A.}~\bibnamefont
  {{Ghosh}}}, \bibinfo {author} {\bibfnamefont {C.}~\bibnamefont {{Kant
  Mishra}}}, \bibinfo {author} {\bibfnamefont {P.}~\bibnamefont {{Ajith}}},
  \bibinfo {author} {\bibfnamefont {W.}~\bibnamefont {{Del Pozzo}}}, \bibinfo
  {author} {\bibfnamefont {C.~P.~L.}\ \bibnamefont {{Berry}}}, \bibinfo
  {author} {\bibfnamefont {A.~B.}\ \bibnamefont {{Nielsen}}}, \ and\ \bibinfo
  {author} {\bibfnamefont {L.}~\bibnamefont {{London}}},\ }\href {\doibase
  10.1088/1361-6382/aa972e} {\bibfield  {journal} {\bibinfo  {journal} {\cqg}\
  }\textbf {\bibinfo {volume} {35}},\ \bibinfo {eid} {014002} (\bibinfo {year}
  {2018})},\ \Eprint {http://arxiv.org/abs/1704.06784} {arXiv:1704.06784
  [gr-qc]} \BibitemShut {NoStop}%
\bibitem [{\citenamefont {{Abbott}}\ \emph
  {et~al.}(2016{\natexlab{a}})\citenamefont {{Abbott}} \emph
  {et~al.}}]{2016PhRvL.116v1101A}%
  \BibitemOpen
  \bibfield  {author} {\bibinfo {author} {\bibfnamefont {B.~P.}\ \bibnamefont
  {{Abbott}}} \emph {et~al.} (\bibinfo {collaboration} {LIGO Scientific
  Collaboration and Virgo Collaboration}),\ }\href {\doibase
  10.1103/PhysRevLett.116.221101} {\bibfield  {journal} {\bibinfo  {journal}
  {\prl}\ }\textbf {\bibinfo {volume} {116}},\ \bibinfo {eid} {221101}
  (\bibinfo {year} {2016}{\natexlab{a}})},\ \Eprint
  {http://arxiv.org/abs/1602.03841} {arXiv:1602.03841 [gr-qc]} \BibitemShut
  {NoStop}%
\bibitem [{\citenamefont {{Pretorius}}(2005)}]{2005PhRvL..95l1101P}%
  \BibitemOpen
  \bibfield  {author} {\bibinfo {author} {\bibfnamefont {F.}~\bibnamefont
  {{Pretorius}}},\ }\href {\doibase 10.1103/PhysRevLett.95.121101} {\bibfield
  {journal} {\bibinfo  {journal} {\prl}\ }\textbf {\bibinfo {volume} {95}},\
  \bibinfo {eid} {121101} (\bibinfo {year} {2005})},\ \Eprint
  {http://arxiv.org/abs/gr-qc/0507014} {gr-qc/0507014} \BibitemShut {NoStop}%
\bibitem [{\citenamefont {{Campanelli}}\ \emph
  {et~al.}(2006{\natexlab{a}})\citenamefont {{Campanelli}}, \citenamefont
  {{Lousto}}, \citenamefont {{Marronetti}},\ and\ \citenamefont
  {{Zlochower}}}]{2006PhRvL..96k1101C}%
  \BibitemOpen
  \bibfield  {author} {\bibinfo {author} {\bibfnamefont {M.}~\bibnamefont
  {{Campanelli}}}, \bibinfo {author} {\bibfnamefont {C.~O.}\ \bibnamefont
  {{Lousto}}}, \bibinfo {author} {\bibfnamefont {P.}~\bibnamefont
  {{Marronetti}}}, \ and\ \bibinfo {author} {\bibfnamefont {Y.}~\bibnamefont
  {{Zlochower}}},\ }\href {\doibase 10.1103/PhysRevLett.96.111101} {\bibfield
  {journal} {\bibinfo  {journal} {\prl}\ }\textbf {\bibinfo {volume} {96}},\
  \bibinfo {eid} {111101} (\bibinfo {year} {2006}{\natexlab{a}})},\ \Eprint
  {http://arxiv.org/abs/gr-qc/0511048} {gr-qc/0511048} \BibitemShut {NoStop}%
\bibitem [{\citenamefont {{Scheel}}\ \emph {et~al.}(2009)\citenamefont
  {{Scheel}}, \citenamefont {{Boyle}}, \citenamefont {{Chu}}, \citenamefont
  {{Kidder}}, \citenamefont {{Matthews}},\ and\ \citenamefont
  {{Pfeiffer}}}]{2009PhRvD..79b4003S}%
  \BibitemOpen
  \bibfield  {author} {\bibinfo {author} {\bibfnamefont {M.~A.}\ \bibnamefont
  {{Scheel}}}, \bibinfo {author} {\bibfnamefont {M.}~\bibnamefont {{Boyle}}},
  \bibinfo {author} {\bibfnamefont {T.}~\bibnamefont {{Chu}}}, \bibinfo
  {author} {\bibfnamefont {L.~E.}\ \bibnamefont {{Kidder}}}, \bibinfo {author}
  {\bibfnamefont {K.~D.}\ \bibnamefont {{Matthews}}}, \ and\ \bibinfo {author}
  {\bibfnamefont {H.~P.}\ \bibnamefont {{Pfeiffer}}},\ }\href {\doibase
  10.1103/PhysRevD.79.024003} {\bibfield  {journal} {\bibinfo  {journal}
  {\prd}\ }\textbf {\bibinfo {volume} {79}},\ \bibinfo {eid} {024003} (\bibinfo
  {year} {2009})},\ \Eprint {http://arxiv.org/abs/0810.1767} {arXiv:0810.1767
  [gr-qc]} \BibitemShut {NoStop}%
\bibitem [{\citenamefont {{Herrmann}}\ \emph {et~al.}(2007)\citenamefont
  {{Herrmann}}, \citenamefont {{Hinder}}, \citenamefont {{Shoemaker}},
  \citenamefont {{Laguna}},\ and\ \citenamefont
  {{Matzner}}}]{2007PhRvD..76h4032H}%
  \BibitemOpen
  \bibfield  {author} {\bibinfo {author} {\bibfnamefont {F.}~\bibnamefont
  {{Herrmann}}}, \bibinfo {author} {\bibfnamefont {I.}~\bibnamefont
  {{Hinder}}}, \bibinfo {author} {\bibfnamefont {D.~M.}\ \bibnamefont
  {{Shoemaker}}}, \bibinfo {author} {\bibfnamefont {P.}~\bibnamefont
  {{Laguna}}}, \ and\ \bibinfo {author} {\bibfnamefont {R.~A.}\ \bibnamefont
  {{Matzner}}},\ }\href {\doibase 10.1103/PhysRevD.76.084032} {\bibfield
  {journal} {\bibinfo  {journal} {\prd}\ }\textbf {\bibinfo {volume} {76}},\
  \bibinfo {eid} {084032} (\bibinfo {year} {2007})},\ \Eprint
  {http://arxiv.org/abs/0706.2541} {arXiv:0706.2541 [gr-qc]} \BibitemShut
  {NoStop}%
\bibitem [{\citenamefont {{Campanelli}}\ \emph
  {et~al.}(2007{\natexlab{a}})\citenamefont {{Campanelli}}, \citenamefont
  {{Lousto}}, \citenamefont {{Zlochower}},\ and\ \citenamefont
  {{Merritt}}}]{2007PhRvL..98w1102C}%
  \BibitemOpen
  \bibfield  {author} {\bibinfo {author} {\bibfnamefont {M.}~\bibnamefont
  {{Campanelli}}}, \bibinfo {author} {\bibfnamefont {C.~O.}\ \bibnamefont
  {{Lousto}}}, \bibinfo {author} {\bibfnamefont {Y.}~\bibnamefont
  {{Zlochower}}}, \ and\ \bibinfo {author} {\bibfnamefont {D.}~\bibnamefont
  {{Merritt}}},\ }\href {\doibase 10.1103/PhysRevLett.98.231102} {\bibfield
  {journal} {\bibinfo  {journal} {\prl}\ }\textbf {\bibinfo {volume} {98}},\
  \bibinfo {eid} {231102} (\bibinfo {year} {2007}{\natexlab{a}})},\ \Eprint
  {http://arxiv.org/abs/gr-qc/0702133} {gr-qc/0702133} \BibitemShut {NoStop}%
\bibitem [{\citenamefont {{Gonz{\'a}lez}}\ \emph
  {et~al.}(2007{\natexlab{a}})\citenamefont {{Gonz{\'a}lez}}, \citenamefont
  {{Hannam}}, \citenamefont {{Sperhake}}, \citenamefont {{Br{\"u}gmann}},\ and\
  \citenamefont {{Husa}}}]{2007PhRvL..98w1101G}%
  \BibitemOpen
  \bibfield  {author} {\bibinfo {author} {\bibfnamefont {J.~A.}\ \bibnamefont
  {{Gonz{\'a}lez}}}, \bibinfo {author} {\bibfnamefont {M.}~\bibnamefont
  {{Hannam}}}, \bibinfo {author} {\bibfnamefont {U.}~\bibnamefont
  {{Sperhake}}}, \bibinfo {author} {\bibfnamefont {B.}~\bibnamefont
  {{Br{\"u}gmann}}}, \ and\ \bibinfo {author} {\bibfnamefont {S.}~\bibnamefont
  {{Husa}}},\ }\href {\doibase 10.1103/PhysRevLett.98.231101} {\bibfield
  {journal} {\bibinfo  {journal} {\prl}\ }\textbf {\bibinfo {volume} {98}},\
  \bibinfo {eid} {231101} (\bibinfo {year} {2007}{\natexlab{a}})},\ \Eprint
  {http://arxiv.org/abs/gr-qc/0702052} {gr-qc/0702052} \BibitemShut {NoStop}%
\bibitem [{\citenamefont {{Gonz{\'a}lez}}\ \emph
  {et~al.}(2007{\natexlab{b}})\citenamefont {{Gonz{\'a}lez}}, \citenamefont
  {{Sperhake}}, \citenamefont {{Br{\"u}gmann}}, \citenamefont {{Hannam}},\ and\
  \citenamefont {{Husa}}}]{2007PhRvL..98i1101G}%
  \BibitemOpen
  \bibfield  {author} {\bibinfo {author} {\bibfnamefont {J.~A.}\ \bibnamefont
  {{Gonz{\'a}lez}}}, \bibinfo {author} {\bibfnamefont {U.}~\bibnamefont
  {{Sperhake}}}, \bibinfo {author} {\bibfnamefont {B.}~\bibnamefont
  {{Br{\"u}gmann}}}, \bibinfo {author} {\bibfnamefont {M.}~\bibnamefont
  {{Hannam}}}, \ and\ \bibinfo {author} {\bibfnamefont {S.}~\bibnamefont
  {{Husa}}},\ }\href {\doibase 10.1103/PhysRevLett.98.091101} {\bibfield
  {journal} {\bibinfo  {journal} {\prl}\ }\textbf {\bibinfo {volume} {98}},\
  \bibinfo {eid} {091101} (\bibinfo {year} {2007}{\natexlab{b}})},\ \Eprint
  {http://arxiv.org/abs/gr-qc/0610154} {gr-qc/0610154} \BibitemShut {NoStop}%
\bibitem [{\citenamefont {{Campanelli}}\ \emph
  {et~al.}(2007{\natexlab{b}})\citenamefont {{Campanelli}}, \citenamefont
  {{Lousto}}, \citenamefont {{Zlochower}},\ and\ \citenamefont
  {{Merritt}}}]{2007ApJ...659L...5C}%
  \BibitemOpen
  \bibfield  {author} {\bibinfo {author} {\bibfnamefont {M.}~\bibnamefont
  {{Campanelli}}}, \bibinfo {author} {\bibfnamefont {C.}~\bibnamefont
  {{Lousto}}}, \bibinfo {author} {\bibfnamefont {Y.}~\bibnamefont
  {{Zlochower}}}, \ and\ \bibinfo {author} {\bibfnamefont {D.}~\bibnamefont
  {{Merritt}}},\ }\href {\doibase 10.1086/516712} {\bibfield  {journal}
  {\bibinfo  {journal} {\apjl}\ }\textbf {\bibinfo {volume} {659}},\ \bibinfo
  {pages} {L5} (\bibinfo {year} {2007}{\natexlab{b}})},\ \Eprint
  {http://arxiv.org/abs/gr-qc/0701164} {gr-qc/0701164} \BibitemShut {NoStop}%
\bibitem [{\citenamefont {{Rezzolla}}\ \emph
  {et~al.}(2008{\natexlab{a}})\citenamefont {{Rezzolla}}, \citenamefont
  {{Barausse}}, \citenamefont {{Dorband}}, \citenamefont {{Pollney}},
  \citenamefont {{Reisswig}}, \citenamefont {{Seiler}},\ and\ \citenamefont
  {{Husa}}}]{2008PhRvD..78d4002R}%
  \BibitemOpen
  \bibfield  {author} {\bibinfo {author} {\bibfnamefont {L.}~\bibnamefont
  {{Rezzolla}}}, \bibinfo {author} {\bibfnamefont {E.}~\bibnamefont
  {{Barausse}}}, \bibinfo {author} {\bibfnamefont {E.~N.}\ \bibnamefont
  {{Dorband}}}, \bibinfo {author} {\bibfnamefont {D.}~\bibnamefont
  {{Pollney}}}, \bibinfo {author} {\bibfnamefont {C.}~\bibnamefont
  {{Reisswig}}}, \bibinfo {author} {\bibfnamefont {J.}~\bibnamefont
  {{Seiler}}}, \ and\ \bibinfo {author} {\bibfnamefont {S.}~\bibnamefont
  {{Husa}}},\ }\href {\doibase 10.1103/PhysRevD.78.044002} {\bibfield
  {journal} {\bibinfo  {journal} {\prd}\ }\textbf {\bibinfo {volume} {78}},\
  \bibinfo {eid} {044002} (\bibinfo {year} {2008}{\natexlab{a}})},\ \Eprint
  {http://arxiv.org/abs/0712.3541} {arXiv:0712.3541 [gr-qc]} \BibitemShut
  {NoStop}%
\bibitem [{\citenamefont {{Rezzolla}}\ \emph
  {et~al.}(2008{\natexlab{b}})\citenamefont {{Rezzolla}}, \citenamefont
  {{Diener}}, \citenamefont {{Dorband}}, \citenamefont {{Pollney}},
  \citenamefont {{Reisswig}}, \citenamefont {{Schnetter}},\ and\ \citenamefont
  {{Seiler}}}]{2008ApJ...674L..29R}%
  \BibitemOpen
  \bibfield  {author} {\bibinfo {author} {\bibfnamefont {L.}~\bibnamefont
  {{Rezzolla}}}, \bibinfo {author} {\bibfnamefont {P.}~\bibnamefont
  {{Diener}}}, \bibinfo {author} {\bibfnamefont {E.~N.}\ \bibnamefont
  {{Dorband}}}, \bibinfo {author} {\bibfnamefont {D.}~\bibnamefont
  {{Pollney}}}, \bibinfo {author} {\bibfnamefont {C.}~\bibnamefont
  {{Reisswig}}}, \bibinfo {author} {\bibfnamefont {E.}~\bibnamefont
  {{Schnetter}}}, \ and\ \bibinfo {author} {\bibfnamefont {J.}~\bibnamefont
  {{Seiler}}},\ }\href {\doibase 10.1086/528935} {\bibfield  {journal}
  {\bibinfo  {journal} {\apjl}\ }\textbf {\bibinfo {volume} {674}},\ \bibinfo
  {pages} {L29} (\bibinfo {year} {2008}{\natexlab{b}})},\ \Eprint
  {http://arxiv.org/abs/0710.3345} {arXiv:0710.3345 [gr-qc]} \BibitemShut
  {NoStop}%
\bibitem [{\citenamefont {{Kesden}}(2008)}]{2008PhRvD..78h4030K}%
  \BibitemOpen
  \bibfield  {author} {\bibinfo {author} {\bibfnamefont {M.}~\bibnamefont
  {{Kesden}}},\ }\href {\doibase 10.1103/PhysRevD.78.084030} {\bibfield
  {journal} {\bibinfo  {journal} {\prd}\ }\textbf {\bibinfo {volume} {78}},\
  \bibinfo {eid} {084030} (\bibinfo {year} {2008})},\ \Eprint
  {http://arxiv.org/abs/0807.3043} {arXiv:0807.3043} \BibitemShut {NoStop}%
\bibitem [{\citenamefont {{Tichy}}\ and\ \citenamefont
  {{Marronetti}}(2008)}]{2008PhRvD..78h1501T}%
  \BibitemOpen
  \bibfield  {author} {\bibinfo {author} {\bibfnamefont {W.}~\bibnamefont
  {{Tichy}}}\ and\ \bibinfo {author} {\bibfnamefont {P.}~\bibnamefont
  {{Marronetti}}},\ }\href {\doibase 10.1103/PhysRevD.78.081501} {\bibfield
  {journal} {\bibinfo  {journal} {\prd}\ }\textbf {\bibinfo {volume} {78}},\
  \bibinfo {eid} {081501} (\bibinfo {year} {2008})},\ \Eprint
  {http://arxiv.org/abs/0807.2985} {arXiv:0807.2985 [gr-qc]} \BibitemShut
  {NoStop}%
\bibitem [{\citenamefont {{Lousto}}\ and\ \citenamefont
  {{Zlochower}}(2008)}]{2008PhRvD..77d4028L}%
  \BibitemOpen
  \bibfield  {author} {\bibinfo {author} {\bibfnamefont {C.~O.}\ \bibnamefont
  {{Lousto}}}\ and\ \bibinfo {author} {\bibfnamefont {Y.}~\bibnamefont
  {{Zlochower}}},\ }\href {\doibase 10.1103/PhysRevD.77.044028} {\bibfield
  {journal} {\bibinfo  {journal} {\prd}\ }\textbf {\bibinfo {volume} {77}},\
  \bibinfo {eid} {044028} (\bibinfo {year} {2008})},\ \Eprint
  {http://arxiv.org/abs/0708.4048} {arXiv:0708.4048 [gr-qc]} \BibitemShut
  {NoStop}%
\bibitem [{\citenamefont {{Barausse}}\ and\ \citenamefont
  {{Rezzolla}}(2009)}]{2009ApJ...704L..40B}%
  \BibitemOpen
  \bibfield  {author} {\bibinfo {author} {\bibfnamefont {E.}~\bibnamefont
  {{Barausse}}}\ and\ \bibinfo {author} {\bibfnamefont {L.}~\bibnamefont
  {{Rezzolla}}},\ }\href {\doibase 10.1088/0004-637X/704/1/L40} {\bibfield
  {journal} {\bibinfo  {journal} {\apjl}\ }\textbf {\bibinfo {volume} {704}},\
  \bibinfo {pages} {L40} (\bibinfo {year} {2009})},\ \Eprint
  {http://arxiv.org/abs/0904.2577} {arXiv:0904.2577 [gr-qc]} \BibitemShut
  {NoStop}%
\bibitem [{\citenamefont {{Pan}}\ \emph {et~al.}(2011)\citenamefont {{Pan}},
  \citenamefont {{Buonanno}}, \citenamefont {{Boyle}}, \citenamefont
  {{Buchman}}, \citenamefont {{Kidder}}, \citenamefont {{Pfeiffer}},\ and\
  \citenamefont {{Scheel}}}]{2011PhRvD..84l4052P}%
  \BibitemOpen
  \bibfield  {author} {\bibinfo {author} {\bibfnamefont {Y.}~\bibnamefont
  {{Pan}}}, \bibinfo {author} {\bibfnamefont {A.}~\bibnamefont {{Buonanno}}},
  \bibinfo {author} {\bibfnamefont {M.}~\bibnamefont {{Boyle}}}, \bibinfo
  {author} {\bibfnamefont {L.~T.}\ \bibnamefont {{Buchman}}}, \bibinfo {author}
  {\bibfnamefont {L.~E.}\ \bibnamefont {{Kidder}}}, \bibinfo {author}
  {\bibfnamefont {H.~P.}\ \bibnamefont {{Pfeiffer}}}, \ and\ \bibinfo {author}
  {\bibfnamefont {M.~A.}\ \bibnamefont {{Scheel}}},\ }\href {\doibase
  10.1103/PhysRevD.84.124052} {\bibfield  {journal} {\bibinfo  {journal}
  {\prd}\ }\textbf {\bibinfo {volume} {84}},\ \bibinfo {eid} {124052} (\bibinfo
  {year} {2011})},\ \Eprint {http://arxiv.org/abs/1106.1021} {arXiv:1106.1021
  [gr-qc]} \BibitemShut {NoStop}%
\bibitem [{\citenamefont {{Barausse}}\ \emph {et~al.}(2012)\citenamefont
  {{Barausse}}, \citenamefont {{Morozova}},\ and\ \citenamefont
  {{Rezzolla}}}]{2012ApJ...758...63B}%
  \BibitemOpen
  \bibfield  {author} {\bibinfo {author} {\bibfnamefont {E.}~\bibnamefont
  {{Barausse}}}, \bibinfo {author} {\bibfnamefont {V.}~\bibnamefont
  {{Morozova}}}, \ and\ \bibinfo {author} {\bibfnamefont {L.}~\bibnamefont
  {{Rezzolla}}},\ }\href {\doibase 10.1088/0004-637X/758/1/63} {\bibfield
  {journal} {\bibinfo  {journal} {\apj}\ }\textbf {\bibinfo {volume} {758}},\
  \bibinfo {eid} {63} (\bibinfo {year} {2012})},\ \bibinfo {note} {[Erratum:
  \apj, 2014, 786, 76]},\ \Eprint {http://arxiv.org/abs/1206.3803}
  {arXiv:1206.3803 [gr-qc]} \BibitemShut {NoStop}%
\bibitem [{\citenamefont {{Lousto}}\ \emph {et~al.}(2012)\citenamefont
  {{Lousto}}, \citenamefont {{Zlochower}}, \citenamefont {{Dotti}},\ and\
  \citenamefont {{Volonteri}}}]{2012PhRvD..85h4015L}%
  \BibitemOpen
  \bibfield  {author} {\bibinfo {author} {\bibfnamefont {C.~O.}\ \bibnamefont
  {{Lousto}}}, \bibinfo {author} {\bibfnamefont {Y.}~\bibnamefont
  {{Zlochower}}}, \bibinfo {author} {\bibfnamefont {M.}~\bibnamefont
  {{Dotti}}}, \ and\ \bibinfo {author} {\bibfnamefont {M.}~\bibnamefont
  {{Volonteri}}},\ }\href {\doibase 10.1103/PhysRevD.85.084015} {\bibfield
  {journal} {\bibinfo  {journal} {\prd}\ }\textbf {\bibinfo {volume} {85}},\
  \bibinfo {eid} {084015} (\bibinfo {year} {2012})},\ \Eprint
  {http://arxiv.org/abs/1201.1923} {arXiv:1201.1923 [gr-qc]} \BibitemShut
  {NoStop}%
\bibitem [{\citenamefont {{Lousto}}\ and\ \citenamefont
  {{Zlochower}}(2013)}]{2013PhRvD..87h4027L}%
  \BibitemOpen
  \bibfield  {author} {\bibinfo {author} {\bibfnamefont {C.~O.}\ \bibnamefont
  {{Lousto}}}\ and\ \bibinfo {author} {\bibfnamefont {Y.}~\bibnamefont
  {{Zlochower}}},\ }\href {\doibase 10.1103/PhysRevD.87.084027} {\bibfield
  {journal} {\bibinfo  {journal} {\prd}\ }\textbf {\bibinfo {volume} {87}},\
  \bibinfo {eid} {084027} (\bibinfo {year} {2013})},\ \Eprint
  {http://arxiv.org/abs/1211.7099} {arXiv:1211.7099 [gr-qc]} \BibitemShut
  {NoStop}%
\bibitem [{\citenamefont {{Healy}}\ \emph {et~al.}(2014)\citenamefont
  {{Healy}}, \citenamefont {{Lousto}},\ and\ \citenamefont
  {{Zlochower}}}]{2014PhRvD..90j4004H}%
  \BibitemOpen
  \bibfield  {author} {\bibinfo {author} {\bibfnamefont {J.}~\bibnamefont
  {{Healy}}}, \bibinfo {author} {\bibfnamefont {C.~O.}\ \bibnamefont
  {{Lousto}}}, \ and\ \bibinfo {author} {\bibfnamefont {Y.}~\bibnamefont
  {{Zlochower}}},\ }\href {\doibase 10.1103/PhysRevD.90.104004} {\bibfield
  {journal} {\bibinfo  {journal} {\prd}\ }\textbf {\bibinfo {volume} {90}},\
  \bibinfo {eid} {104004} (\bibinfo {year} {2014})},\ \Eprint
  {http://arxiv.org/abs/1406.7295} {arXiv:1406.7295 [gr-qc]} \BibitemShut
  {NoStop}%
\bibitem [{\citenamefont {{Zlochower}}\ and\ \citenamefont
  {{Lousto}}(2015)}]{2015PhRvD..92b4022Z}%
  \BibitemOpen
  \bibfield  {author} {\bibinfo {author} {\bibfnamefont {Y.}~\bibnamefont
  {{Zlochower}}}\ and\ \bibinfo {author} {\bibfnamefont {C.~O.}\ \bibnamefont
  {{Lousto}}},\ }\href {\doibase 10.1103/PhysRevD.92.024022} {\bibfield
  {journal} {\bibinfo  {journal} {\prd}\ }\textbf {\bibinfo {volume} {92}},\
  \bibinfo {eid} {024022} (\bibinfo {year} {2015})},\ \Eprint
  {http://arxiv.org/abs/1503.07536} {arXiv:1503.07536 [gr-qc]} \BibitemShut
  {NoStop}%
\bibitem [{\citenamefont {{Hofmann}}\ \emph {et~al.}(2016)\citenamefont
  {{Hofmann}}, \citenamefont {{Barausse}},\ and\ \citenamefont
  {{Rezzolla}}}]{2016ApJ...825L..19H}%
  \BibitemOpen
  \bibfield  {author} {\bibinfo {author} {\bibfnamefont {F.}~\bibnamefont
  {{Hofmann}}}, \bibinfo {author} {\bibfnamefont {E.}~\bibnamefont
  {{Barausse}}}, \ and\ \bibinfo {author} {\bibfnamefont {L.}~\bibnamefont
  {{Rezzolla}}},\ }\href {\doibase 10.3847/2041-8205/825/2/L19} {\bibfield
  {journal} {\bibinfo  {journal} {\apjl}\ }\textbf {\bibinfo {volume} {825}},\
  \bibinfo {eid} {L19} (\bibinfo {year} {2016})},\ \Eprint
  {http://arxiv.org/abs/1605.01938} {arXiv:1605.01938 [gr-qc]} \BibitemShut
  {NoStop}%
\bibitem [{\citenamefont {{Gerosa}}\ and\ \citenamefont
  {{Kesden}}(2016)}]{2016PhRvD..93l4066G}%
  \BibitemOpen
  \bibfield  {author} {\bibinfo {author} {\bibfnamefont {D.}~\bibnamefont
  {{Gerosa}}}\ and\ \bibinfo {author} {\bibfnamefont {M.}~\bibnamefont
  {{Kesden}}},\ }\href {\doibase 10.1103/PhysRevD.93.124066} {\bibfield
  {journal} {\bibinfo  {journal} {\prd}\ }\textbf {\bibinfo {volume} {93}},\
  \bibinfo {eid} {124066} (\bibinfo {year} {2016})},\ \Eprint
  {http://arxiv.org/abs/1605.01067} {arXiv:1605.01067 [astro-ph.HE]}
  \BibitemShut {NoStop}%
\bibitem [{\citenamefont {{Jim{\'e}nez-Forteza}}\ \emph
  {et~al.}(2017)\citenamefont {{Jim{\'e}nez-Forteza}}, \citenamefont
  {{Keitel}}, \citenamefont {{Husa}}, \citenamefont {{Hannam}}, \citenamefont
  {{Khan}},\ and\ \citenamefont {{P{\"u}rrer}}}]{2017PhRvD..95f4024J}%
  \BibitemOpen
  \bibfield  {author} {\bibinfo {author} {\bibfnamefont {X.}~\bibnamefont
  {{Jim{\'e}nez-Forteza}}}, \bibinfo {author} {\bibfnamefont {D.}~\bibnamefont
  {{Keitel}}}, \bibinfo {author} {\bibfnamefont {S.}~\bibnamefont {{Husa}}},
  \bibinfo {author} {\bibfnamefont {M.}~\bibnamefont {{Hannam}}}, \bibinfo
  {author} {\bibfnamefont {S.}~\bibnamefont {{Khan}}}, \ and\ \bibinfo {author}
  {\bibfnamefont {M.}~\bibnamefont {{P{\"u}rrer}}},\ }\href {\doibase
  10.1103/PhysRevD.95.064024} {\bibfield  {journal} {\bibinfo  {journal}
  {\prd}\ }\textbf {\bibinfo {volume} {95}},\ \bibinfo {eid} {064024} (\bibinfo
  {year} {2017})},\ \Eprint {http://arxiv.org/abs/1611.00332} {arXiv:1611.00332
  [gr-qc]} \BibitemShut {NoStop}%
\bibitem [{\citenamefont {{Healy}}\ and\ \citenamefont
  {{Lousto}}(2017)}]{2017PhRvD..95b4037H}%
  \BibitemOpen
  \bibfield  {author} {\bibinfo {author} {\bibfnamefont {J.}~\bibnamefont
  {{Healy}}}\ and\ \bibinfo {author} {\bibfnamefont {C.~O.}\ \bibnamefont
  {{Lousto}}},\ }\href {\doibase 10.1103/PhysRevD.95.024037} {\bibfield
  {journal} {\bibinfo  {journal} {\prd}\ }\textbf {\bibinfo {volume} {95}},\
  \bibinfo {eid} {024037} (\bibinfo {year} {2017})},\ \Eprint
  {http://arxiv.org/abs/1610.09713} {arXiv:1610.09713 [gr-qc]} \BibitemShut
  {NoStop}%
\bibitem [{\citenamefont {{Healy}}\ and\ \citenamefont
  {{Lousto}}(2018)}]{2018PhRvD..97h4002H}%
  \BibitemOpen
  \bibfield  {author} {\bibinfo {author} {\bibfnamefont {J.}~\bibnamefont
  {{Healy}}}\ and\ \bibinfo {author} {\bibfnamefont {C.~O.}\ \bibnamefont
  {{Lousto}}},\ }\href {\doibase 10.1103/PhysRevD.97.084002} {\bibfield
  {journal} {\bibinfo  {journal} {\prd}\ }\textbf {\bibinfo {volume} {97}},\
  \bibinfo {eid} {084002} (\bibinfo {year} {2018})},\ \Eprint
  {http://arxiv.org/abs/1801.08162} {arXiv:1801.08162 [gr-qc]} \BibitemShut
  {NoStop}%
\bibitem [{\citenamefont {{Abbott}}\ \emph
  {et~al.}(2016{\natexlab{b}})\citenamefont {{Abbott}} \emph
  {et~al.}}]{2016PhRvX...6d1015A}%
  \BibitemOpen
  \bibfield  {author} {\bibinfo {author} {\bibfnamefont {B.~P.}\ \bibnamefont
  {{Abbott}}} \emph {et~al.} (\bibinfo {collaboration} {LIGO Scientific
  Collaboration and Virgo Collaboration}),\ }\href {\doibase
  10.1103/PhysRevX.6.041015} {\bibfield  {journal} {\bibinfo  {journal} {\prx}\
  }\textbf {\bibinfo {volume} {6}},\ \bibinfo {eid} {041015} (\bibinfo {year}
  {2016}{\natexlab{b}})},\ \Eprint {http://arxiv.org/abs/1606.04856}
  {arXiv:1606.04856 [gr-qc]} \BibitemShut {NoStop}%
\bibitem [{\citenamefont {{Abbott}}\ \emph
  {et~al.}(2017{\natexlab{a}})\citenamefont {{Abbott}} \emph
  {et~al.}}]{2017PhRvL.118v1101A}%
  \BibitemOpen
  \bibfield  {author} {\bibinfo {author} {\bibfnamefont {B.~P.}\ \bibnamefont
  {{Abbott}}} \emph {et~al.} (\bibinfo {collaboration} {LIGO Scientific
  Collaboration and Virgo Collaboration}),\ }\href {\doibase
  10.1103/PhysRevLett.118.221101} {\bibfield  {journal} {\bibinfo  {journal}
  {\prl}\ }\textbf {\bibinfo {volume} {118}},\ \bibinfo {eid} {221101}
  (\bibinfo {year} {2017}{\natexlab{a}})},\ \bibinfo {note} {[Erratum \prl,
  2018, 21, 129901]},\ \Eprint {http://arxiv.org/abs/1706.01812}
  {arXiv:1706.01812 [gr-qc]} \BibitemShut {NoStop}%
\bibitem [{\citenamefont {{Johnson-McDaniel}}\ \emph {et~al.}()\citenamefont
  {{Johnson-McDaniel}}, \citenamefont {{Gupta}}, \citenamefont {{Ajith}},
  \citenamefont {{Keitel}}, \citenamefont {{Birnholtz}}, \citenamefont
  {{Ohme}},\ and\ \citenamefont {{Husa}}}]{dccprec}%
  \BibitemOpen
  \bibfield  {author} {\bibinfo {author} {\bibfnamefont {N.~K.}\ \bibnamefont
  {{Johnson-McDaniel}}}, \bibinfo {author} {\bibfnamefont {A.}~\bibnamefont
  {{Gupta}}}, \bibinfo {author} {\bibfnamefont {P.}~\bibnamefont {{Ajith}}},
  \bibinfo {author} {\bibfnamefont {D.}~\bibnamefont {{Keitel}}}, \bibinfo
  {author} {\bibfnamefont {O.}~\bibnamefont {{Birnholtz}}}, \bibinfo {author}
  {\bibfnamefont {F.}~\bibnamefont {{Ohme}}}, \ and\ \bibinfo {author}
  {\bibfnamefont {S.}~\bibnamefont {{Husa}}},\ }\href@noop {} {\ }\bibinfo
  {note}
  {\href{https://dcc.ligo.org/T1600168/public}{dcc.ligo.org/T1600168/public}}\BibitemShut
  {NoStop}%
\bibitem [{\citenamefont {{Kesden}}\ \emph {et~al.}(2010)\citenamefont
  {{Kesden}}, \citenamefont {{Sperhake}},\ and\ \citenamefont
  {{Berti}}}]{2010PhRvD..81h4054K}%
  \BibitemOpen
  \bibfield  {author} {\bibinfo {author} {\bibfnamefont {M.}~\bibnamefont
  {{Kesden}}}, \bibinfo {author} {\bibfnamefont {U.}~\bibnamefont
  {{Sperhake}}}, \ and\ \bibinfo {author} {\bibfnamefont {E.}~\bibnamefont
  {{Berti}}},\ }\href {\doibase 10.1103/PhysRevD.81.084054} {\bibfield
  {journal} {\bibinfo  {journal} {\prd}\ }\textbf {\bibinfo {volume} {81}},\
  \bibinfo {eid} {084054} (\bibinfo {year} {2010})},\ \Eprint
  {http://arxiv.org/abs/1002.2643} {arXiv:1002.2643 [astro-ph.GA]} \BibitemShut
  {NoStop}%
\bibitem [{\citenamefont {{Kidder}}\ \emph {et~al.}(2000)\citenamefont
  {{Kidder}}, \citenamefont {{Scheel}}, \citenamefont {{Teukolsky}},
  \citenamefont {{Carlson}},\ and\ \citenamefont
  {{Cook}}}]{2000PhRvD..62h4032K}%
  \BibitemOpen
  \bibfield  {author} {\bibinfo {author} {\bibfnamefont {L.~E.}\ \bibnamefont
  {{Kidder}}}, \bibinfo {author} {\bibfnamefont {M.~A.}\ \bibnamefont
  {{Scheel}}}, \bibinfo {author} {\bibfnamefont {S.~A.}\ \bibnamefont
  {{Teukolsky}}}, \bibinfo {author} {\bibfnamefont {E.~D.}\ \bibnamefont
  {{Carlson}}}, \ and\ \bibinfo {author} {\bibfnamefont {G.~B.}\ \bibnamefont
  {{Cook}}},\ }\href {\doibase 10.1103/PhysRevD.62.084032} {\bibfield
  {journal} {\bibinfo  {journal} {\prd}\ }\textbf {\bibinfo {volume} {62}},\
  \bibinfo {eid} {084032} (\bibinfo {year} {2000})},\ \Eprint
  {http://arxiv.org/abs/gr-qc/0005056} {gr-qc/0005056} \BibitemShut {NoStop}%
\bibitem [{\citenamefont {{Varma}}\ \emph {et~al.}()\citenamefont {{Varma}}
  \emph {et~al.}}]{surfinBH}%
  \BibitemOpen
  \bibfield  {author} {\bibinfo {author} {\bibfnamefont {V.}~\bibnamefont
  {{Varma}}} \emph {et~al.},\ }\href@noop {} {\ }\bibinfo {note}
  {\href{https://pypi.org/project/surfinBH}{pypi.org/project/surfinBH},
  \href{https://doi.org/10.5281/zenodo.1418525}{doi.org/10.5281/zenodo.1418525}}\BibitemShut
  {NoStop}%
\bibitem [{\citenamefont {{Rasmussen}}\ and\ \citenamefont
  {{Williams}}(2006)}]{2006gpml.book.....R}%
  \BibitemOpen
  \bibfield  {author} {\bibinfo {author} {\bibfnamefont {C.~E.}\ \bibnamefont
  {{Rasmussen}}}\ and\ \bibinfo {author} {\bibfnamefont {C.~K.~I.}\
  \bibnamefont {{Williams}}},\ }\href@noop {} {\emph {\bibinfo {title}
  {Gaussian Processes for Machine Learning, by C.E.~Rasmussen and
  C.K.I.~Williams.~ISBN-13 978-0-262-18253-9}}}\ (\bibinfo {year}
  {2006})\BibitemShut {NoStop}%
\bibitem [{sur()}]{surfsupplement}%
  \BibitemOpen
  \href@noop {} {}\bibinfo {howpublished} {See Supplemental Material
  \hyperlink{page.8}{here}, for details of the GPR fitting method, more
  detailed exploration of extrapolation errors, tests on the efficacy of GPR's
  error prediction, and details of the public Python implementation. This
  further includes Refs.~\cite{ 1994PhRvD..49.2658C, 1995PhRvD..52..848P,
  2003itil.book.....M, 2011PhRvD..84h4037A, 2012arXiv1201.0490P,
  2013PhRvL.111x1104M, 2015PhRvD..92j2001K, 2015PhRvL.115l1102B,
  2016CQGra..33p5001C, Jones:2001aa, Walt, h5py, travis-ci}.}\BibitemShut
  {Stop}%
\bibitem [{\citenamefont {Varma}\ \emph {et~al.}(2018)\citenamefont {Varma},
  \citenamefont {Field}, \citenamefont {Scheel}, \citenamefont {Blackman},
  \citenamefont {Kidder},\ and\ \citenamefont {Pfeiffer}}]{Varma:2018hybsur}%
  \BibitemOpen
  \bibfield  {author} {\bibinfo {author} {\bibfnamefont {V.}~\bibnamefont
  {Varma}}, \bibinfo {author} {\bibfnamefont {S.}~\bibnamefont {Field}},
  \bibinfo {author} {\bibfnamefont {M.~A.}\ \bibnamefont {Scheel}}, \bibinfo
  {author} {\bibfnamefont {J.}~\bibnamefont {Blackman}}, \bibinfo {author}
  {\bibfnamefont {L.~E.}\ \bibnamefont {Kidder}}, \ and\ \bibinfo {author}
  {\bibfnamefont {H.~P.}\ \bibnamefont {Pfeiffer}},\ }\href@noop {} {\
  (\bibinfo {year} {2018})},\ \Eprint {http://arxiv.org/abs/1812.07865}
  {arXiv:1812.07865 [gr-qc]} \BibitemShut {NoStop}%
\bibitem [{\citenamefont {{Lovelace}}\ \emph {et~al.}(2008)\citenamefont
  {{Lovelace}}, \citenamefont {{Owen}}, \citenamefont {{Pfeiffer}},\ and\
  \citenamefont {{Chu}}}]{2008PhRvD..78h4017L}%
  \BibitemOpen
  \bibfield  {author} {\bibinfo {author} {\bibfnamefont {G.}~\bibnamefont
  {{Lovelace}}}, \bibinfo {author} {\bibfnamefont {R.}~\bibnamefont {{Owen}}},
  \bibinfo {author} {\bibfnamefont {H.~P.}\ \bibnamefont {{Pfeiffer}}}, \ and\
  \bibinfo {author} {\bibfnamefont {T.}~\bibnamefont {{Chu}}},\ }\href
  {\doibase 10.1103/PhysRevD.78.084017} {\bibfield  {journal} {\bibinfo
  {journal} {\prd}\ }\textbf {\bibinfo {volume} {78}},\ \bibinfo {eid} {084017}
  (\bibinfo {year} {2008})},\ \Eprint {http://arxiv.org/abs/0805.4192}
  {arXiv:0805.4192 [gr-qc]} \BibitemShut {NoStop}%
\bibitem [{\citenamefont {{Boyle}}\ \emph {et~al.}(2018)\citenamefont {{Boyle}}
  \emph {et~al.}}]{Catalog2018}%
  \BibitemOpen
  \bibfield  {author} {\bibinfo {author} {\bibfnamefont {M.}~\bibnamefont
  {{Boyle}}} \emph {et~al.},\ }\href@noop {} {\  (\bibinfo {year} {2018})},\
  \bibinfo {note} {in preparation}\BibitemShut {NoStop}%
\bibitem [{\citenamefont {{Gerosa}}\ \emph {et~al.}(2018)\citenamefont
  {{Gerosa}}, \citenamefont {{H{\'e}bert}},\ and\ \citenamefont
  {{Stein}}}]{2018PhRvD..97j4049G}%
  \BibitemOpen
  \bibfield  {author} {\bibinfo {author} {\bibfnamefont {D.}~\bibnamefont
  {{Gerosa}}}, \bibinfo {author} {\bibfnamefont {F.}~\bibnamefont
  {{H{\'e}bert}}}, \ and\ \bibinfo {author} {\bibfnamefont {L.~C.}\
  \bibnamefont {{Stein}}},\ }\href {\doibase 10.1103/PhysRevD.97.104049}
  {\bibfield  {journal} {\bibinfo  {journal} {\prd}\ }\textbf {\bibinfo
  {volume} {97}},\ \bibinfo {eid} {104049} (\bibinfo {year} {2018})},\ \Eprint
  {http://arxiv.org/abs/1802.04276} {arXiv:1802.04276 [gr-qc]} \BibitemShut
  {NoStop}%
\bibitem [{\citenamefont {{Ruiz}}\ \emph {et~al.}(2008)\citenamefont {{Ruiz}},
  \citenamefont {{Alcubierre}}, \citenamefont {{N{\'u}{\~n}ez}},\ and\
  \citenamefont {{Takahashi}}}]{2008GReGr..40.1705R}%
  \BibitemOpen
  \bibfield  {author} {\bibinfo {author} {\bibfnamefont {M.}~\bibnamefont
  {{Ruiz}}}, \bibinfo {author} {\bibfnamefont {M.}~\bibnamefont
  {{Alcubierre}}}, \bibinfo {author} {\bibfnamefont {D.}~\bibnamefont
  {{N{\'u}{\~n}ez}}}, \ and\ \bibinfo {author} {\bibfnamefont {R.}~\bibnamefont
  {{Takahashi}}},\ }\href {\doibase 10.1007/s10714-007-0570-8} {\bibfield
  {journal} {\bibinfo  {journal} {General Relativity and Gravitation}\ }\textbf
  {\bibinfo {volume} {40}},\ \bibinfo {pages} {1705} (\bibinfo {year}
  {2008})},\ \Eprint {http://arxiv.org/abs/0707.4654} {arXiv:0707.4654 [gr-qc]}
  \BibitemShut {NoStop}%
\bibitem [{\citenamefont {{Boyle}}\ and\ \citenamefont
  {{Mrou{\'e}}}(2009)}]{2009PhRvD..80l4045B}%
  \BibitemOpen
  \bibfield  {author} {\bibinfo {author} {\bibfnamefont {M.}~\bibnamefont
  {{Boyle}}}\ and\ \bibinfo {author} {\bibfnamefont {A.~H.}\ \bibnamefont
  {{Mrou{\'e}}}},\ }\href {\doibase 10.1103/PhysRevD.80.124045} {\bibfield
  {journal} {\bibinfo  {journal} {\prd}\ }\textbf {\bibinfo {volume} {80}},\
  \bibinfo {eid} {124045} (\bibinfo {year} {2009})},\ \Eprint
  {http://arxiv.org/abs/0905.3177} {arXiv:0905.3177 [gr-qc]} \BibitemShut
  {NoStop}%
\bibitem [{\citenamefont {{Boyle}}\ \emph {et~al.}(2008)\citenamefont
  {{Boyle}}, \citenamefont {{Kesden}},\ and\ \citenamefont
  {{Nissanke}}}]{2008PhRvL.100o1101B}%
  \BibitemOpen
  \bibfield  {author} {\bibinfo {author} {\bibfnamefont {L.}~\bibnamefont
  {{Boyle}}}, \bibinfo {author} {\bibfnamefont {M.}~\bibnamefont {{Kesden}}}, \
  and\ \bibinfo {author} {\bibfnamefont {S.}~\bibnamefont {{Nissanke}}},\
  }\href {\doibase 10.1103/PhysRevLett.100.151101} {\bibfield  {journal}
  {\bibinfo  {journal} {\prl}\ }\textbf {\bibinfo {volume} {100}},\ \bibinfo
  {eid} {151101} (\bibinfo {year} {2008})},\ \Eprint
  {http://arxiv.org/abs/0709.0299} {arXiv:0709.0299 [gr-qc]} \BibitemShut
  {NoStop}%
\bibitem [{\citenamefont {{Berti}}\ \emph {et~al.}(2012)\citenamefont
  {{Berti}}, \citenamefont {{Kesden}},\ and\ \citenamefont
  {{Sperhake}}}]{2012PhRvD..85l4049B}%
  \BibitemOpen
  \bibfield  {author} {\bibinfo {author} {\bibfnamefont {E.}~\bibnamefont
  {{Berti}}}, \bibinfo {author} {\bibfnamefont {M.}~\bibnamefont {{Kesden}}}, \
  and\ \bibinfo {author} {\bibfnamefont {U.}~\bibnamefont {{Sperhake}}},\
  }\href {\doibase 10.1103/PhysRevD.85.124049} {\bibfield  {journal} {\bibinfo
  {journal} {\prd}\ }\textbf {\bibinfo {volume} {85}},\ \bibinfo {eid} {124049}
  (\bibinfo {year} {2012})},\ \Eprint {http://arxiv.org/abs/1203.2920}
  {arXiv:1203.2920 [astro-ph.HE]} \BibitemShut {NoStop}%
\bibitem [{\citenamefont {{Campanelli}}\ \emph
  {et~al.}(2006{\natexlab{b}})\citenamefont {{Campanelli}}, \citenamefont
  {{Lousto}},\ and\ \citenamefont {{Zlochower}}}]{2006PhRvD..74d1501C}%
  \BibitemOpen
  \bibfield  {author} {\bibinfo {author} {\bibfnamefont {M.}~\bibnamefont
  {{Campanelli}}}, \bibinfo {author} {\bibfnamefont {C.~O.}\ \bibnamefont
  {{Lousto}}}, \ and\ \bibinfo {author} {\bibfnamefont {Y.}~\bibnamefont
  {{Zlochower}}},\ }\href {\doibase 10.1103/PhysRevD.74.041501} {\bibfield
  {journal} {\bibinfo  {journal} {\prd}\ }\textbf {\bibinfo {volume} {74}},\
  \bibinfo {eid} {041501} (\bibinfo {year} {2006}{\natexlab{b}})},\ \Eprint
  {http://arxiv.org/abs/gr-qc/0604012} {gr-qc/0604012} \BibitemShut {NoStop}%
\bibitem [{\citenamefont {{Lousto}}\ and\ \citenamefont
  {{Zlochower}}(2014)}]{2014PhRvD..89b1501L}%
  \BibitemOpen
  \bibfield  {author} {\bibinfo {author} {\bibfnamefont {C.~O.}\ \bibnamefont
  {{Lousto}}}\ and\ \bibinfo {author} {\bibfnamefont {Y.}~\bibnamefont
  {{Zlochower}}},\ }\href {\doibase 10.1103/PhysRevD.89.021501} {\bibfield
  {journal} {\bibinfo  {journal} {\prd}\ }\textbf {\bibinfo {volume} {89}},\
  \bibinfo {eid} {021501} (\bibinfo {year} {2014})},\ \Eprint
  {http://arxiv.org/abs/1307.6237} {arXiv:1307.6237 [gr-qc]} \BibitemShut
  {NoStop}%
\bibitem [{\citenamefont {{Scheel}}\ \emph {et~al.}(2015)\citenamefont
  {{Scheel}}, \citenamefont {{Giesler}}, \citenamefont {{Hemberger}},
  \citenamefont {{Lovelace}}, \citenamefont {{Kuper}}, \citenamefont {{Boyle}},
  \citenamefont {{Szil{\'a}gyi}},\ and\ \citenamefont
  {{Kidder}}}]{2015CQGra..32j5009S}%
  \BibitemOpen
  \bibfield  {author} {\bibinfo {author} {\bibfnamefont {M.~A.}\ \bibnamefont
  {{Scheel}}}, \bibinfo {author} {\bibfnamefont {M.}~\bibnamefont {{Giesler}}},
  \bibinfo {author} {\bibfnamefont {D.~A.}\ \bibnamefont {{Hemberger}}},
  \bibinfo {author} {\bibfnamefont {G.}~\bibnamefont {{Lovelace}}}, \bibinfo
  {author} {\bibfnamefont {K.}~\bibnamefont {{Kuper}}}, \bibinfo {author}
  {\bibfnamefont {M.}~\bibnamefont {{Boyle}}}, \bibinfo {author} {\bibfnamefont
  {B.}~\bibnamefont {{Szil{\'a}gyi}}}, \ and\ \bibinfo {author} {\bibfnamefont
  {L.~E.}\ \bibnamefont {{Kidder}}},\ }\href {\doibase
  10.1088/0264-9381/32/10/105009} {\bibfield  {journal} {\bibinfo  {journal}
  {\cqg}\ }\textbf {\bibinfo {volume} {32}},\ \bibinfo {eid} {105009} (\bibinfo
  {year} {2015})},\ \Eprint {http://arxiv.org/abs/1412.1803} {arXiv:1412.1803
  [gr-qc]} \BibitemShut {NoStop}%
\bibitem [{\citenamefont {{Buonanno}}\ \emph {et~al.}(2003)\citenamefont
  {{Buonanno}}, \citenamefont {{Chen}},\ and\ \citenamefont
  {{Vallisneri}}}]{2003PhRvD..67j4025B}%
  \BibitemOpen
  \bibfield  {author} {\bibinfo {author} {\bibfnamefont {A.}~\bibnamefont
  {{Buonanno}}}, \bibinfo {author} {\bibfnamefont {Y.}~\bibnamefont {{Chen}}},
  \ and\ \bibinfo {author} {\bibfnamefont {M.}~\bibnamefont {{Vallisneri}}},\
  }\href {\doibase 10.1103/PhysRevD.67.104025} {\bibfield  {journal} {\bibinfo
  {journal} {\prd}\ }\textbf {\bibinfo {volume} {67}},\ \bibinfo {eid} {104025}
  (\bibinfo {year} {2003})},\ \Eprint {http://arxiv.org/abs/gr-qc/0211087}
  {gr-qc/0211087} \BibitemShut {NoStop}%
\bibitem [{\citenamefont {{Boyle}}\ \emph {et~al.}(2007)\citenamefont
  {{Boyle}}, \citenamefont {{Brown}}, \citenamefont {{Kidder}}, \citenamefont
  {{Mrou{\'e}}}, \citenamefont {{Pfeiffer}}, \citenamefont {{Scheel}},
  \citenamefont {{Cook}},\ and\ \citenamefont
  {{Teukolsky}}}]{2007PhRvD..76l4038B}%
  \BibitemOpen
  \bibfield  {author} {\bibinfo {author} {\bibfnamefont {M.}~\bibnamefont
  {{Boyle}}}, \bibinfo {author} {\bibfnamefont {D.~A.}\ \bibnamefont
  {{Brown}}}, \bibinfo {author} {\bibfnamefont {L.~E.}\ \bibnamefont
  {{Kidder}}}, \bibinfo {author} {\bibfnamefont {A.~H.}\ \bibnamefont
  {{Mrou{\'e}}}}, \bibinfo {author} {\bibfnamefont {H.~P.}\ \bibnamefont
  {{Pfeiffer}}}, \bibinfo {author} {\bibfnamefont {M.~A.}\ \bibnamefont
  {{Scheel}}}, \bibinfo {author} {\bibfnamefont {G.~B.}\ \bibnamefont
  {{Cook}}}, \ and\ \bibinfo {author} {\bibfnamefont {S.~A.}\ \bibnamefont
  {{Teukolsky}}},\ }\href {\doibase 10.1103/PhysRevD.76.124038} {\bibfield
  {journal} {\bibinfo  {journal} {\prd}\ }\textbf {\bibinfo {volume} {76}},\
  \bibinfo {eid} {124038} (\bibinfo {year} {2007})},\ \Eprint
  {http://arxiv.org/abs/0710.0158} {arXiv:0710.0158 [gr-qc]} \BibitemShut
  {NoStop}%
\bibitem [{\citenamefont {{Ossokine}}\ \emph {et~al.}(2015)\citenamefont
  {{Ossokine}}, \citenamefont {{Boyle}}, \citenamefont {{Kidder}},
  \citenamefont {{Pfeiffer}}, \citenamefont {{Scheel}},\ and\ \citenamefont
  {{Szil{\'a}gyi}}}]{2015PhRvD..92j4028O}%
  \BibitemOpen
  \bibfield  {author} {\bibinfo {author} {\bibfnamefont {S.}~\bibnamefont
  {{Ossokine}}}, \bibinfo {author} {\bibfnamefont {M.}~\bibnamefont {{Boyle}}},
  \bibinfo {author} {\bibfnamefont {L.~E.}\ \bibnamefont {{Kidder}}}, \bibinfo
  {author} {\bibfnamefont {H.~P.}\ \bibnamefont {{Pfeiffer}}}, \bibinfo
  {author} {\bibfnamefont {M.~A.}\ \bibnamefont {{Scheel}}}, \ and\ \bibinfo
  {author} {\bibfnamefont {B.}~\bibnamefont {{Szil{\'a}gyi}}},\ }\href
  {\doibase 10.1103/PhysRevD.92.104028} {\bibfield  {journal} {\bibinfo
  {journal} {\prd}\ }\textbf {\bibinfo {volume} {92}},\ \bibinfo {eid} {104028}
  (\bibinfo {year} {2015})},\ \Eprint {http://arxiv.org/abs/1502.01747}
  {arXiv:1502.01747 [gr-qc]} \BibitemShut {NoStop}%
\bibitem [{\citenamefont {{Varma}}\ \emph {et~al.}(2019)\citenamefont
  {{Varma}}, \citenamefont {{Field}}, \citenamefont {{Scheel}} \emph
  {et~al.}}]{Inprep-Varma:2019}%
  \BibitemOpen
  \bibfield  {author} {\bibinfo {author} {\bibfnamefont {V.}~\bibnamefont
  {{Varma}}}, \bibinfo {author} {\bibfnamefont {S.}~\bibnamefont {{Field}}},
  \bibinfo {author} {\bibfnamefont {M.~A.}\ \bibnamefont {{Scheel}}},  \emph
  {et~al.},\ }\href@noop {} {\  (\bibinfo {year} {2019})},\ \bibinfo {note} {in
  preparation}\BibitemShut {NoStop}%
\bibitem [{\citenamefont {{Vallisneri}}(2008)}]{2008PhRvD..77d2001V}%
  \BibitemOpen
  \bibfield  {author} {\bibinfo {author} {\bibfnamefont {M.}~\bibnamefont
  {{Vallisneri}}},\ }\href {\doibase 10.1103/PhysRevD.77.042001} {\bibfield
  {journal} {\bibinfo  {journal} {\prd}\ }\textbf {\bibinfo {volume} {77}},\
  \bibinfo {eid} {042001} (\bibinfo {year} {2008})},\ \Eprint
  {http://arxiv.org/abs/gr-qc/0703086} {gr-qc/0703086} \BibitemShut {NoStop}%
\bibitem [{\citenamefont {{Abbott}}\ \emph {et~al.}(2018)\citenamefont
  {{Abbott}} \emph {et~al.}}]{2018LRR....21....3A}%
  \BibitemOpen
  \bibfield  {author} {\bibinfo {author} {\bibfnamefont {B.~P.}\ \bibnamefont
  {{Abbott}}} \emph {et~al.} (\bibinfo {collaboration} {VIRGO, KAGRA, LIGO
  Scientific}),\ }\href {\doibase 10.1007/s41114-018-0012-9} {\bibfield
  {journal} {\bibinfo  {journal} {\lrr}\ }\textbf {\bibinfo {volume} {21}},\
  \bibinfo {eid} {3} (\bibinfo {year} {2018})},\ \Eprint
  {http://arxiv.org/abs/1304.0670} {arXiv:1304.0670 [gr-qc]} \BibitemShut
  {NoStop}%
\bibitem [{\citenamefont {{Amaro-Seoane}}\ \emph {et~al.}(2017)\citenamefont
  {{Amaro-Seoane}} \emph {et~al.}}]{2017arXiv170200786A}%
  \BibitemOpen
  \bibfield  {author} {\bibinfo {author} {\bibfnamefont {P.}~\bibnamefont
  {{Amaro-Seoane}}} \emph {et~al.} (\bibinfo {collaboration} {LISA Core
  Team}),\ }\href@noop {} {\  (\bibinfo {year} {2017})},\ \Eprint
  {http://arxiv.org/abs/1702.00786} {arXiv:1702.00786 [astro-ph.IM]}
  \BibitemShut {NoStop}%
\bibitem [{\citenamefont {{Punturo}}\ \emph {et~al.}(2010)\citenamefont
  {{Punturo}} \emph {et~al.}}]{2010CQGra..27s4002P}%
  \BibitemOpen
  \bibfield  {author} {\bibinfo {author} {\bibfnamefont {M.}~\bibnamefont
  {{Punturo}}} \emph {et~al.},\ }\href {\doibase
  10.1088/0264-9381/27/19/194002} {\bibfield  {journal} {\bibinfo  {journal}
  {\cqg}\ }\textbf {\bibinfo {volume} {27}},\ \bibinfo {eid} {194002} (\bibinfo
  {year} {2010})}\BibitemShut {NoStop}%
\bibitem [{\citenamefont {{Abbott}}\ \emph
  {et~al.}(2017{\natexlab{b}})\citenamefont {{Abbott}} \emph
  {et~al.}}]{2017CQGra..34d4001A}%
  \BibitemOpen
  \bibfield  {author} {\bibinfo {author} {\bibfnamefont {B.~P.}\ \bibnamefont
  {{Abbott}}} \emph {et~al.} (\bibinfo {collaboration} {LIGO Scientific
  Collaboration and Virgo Collaboration}),\ }\href {\doibase
  10.1088/1361-6382/aa51f4} {\bibfield  {journal} {\bibinfo  {journal} {\cqg}\
  }\textbf {\bibinfo {volume} {34}},\ \bibinfo {eid} {044001} (\bibinfo {year}
  {2017}{\natexlab{b}})},\ \Eprint {http://arxiv.org/abs/1607.08697}
  {arXiv:1607.08697 [astro-ph.IM]} \BibitemShut {NoStop}%
\bibitem [{\citenamefont {{Cahillane}}\ \emph {et~al.}(2017)\citenamefont
  {{Cahillane}}, \citenamefont {{Betzwieser}}, \citenamefont {{Brown}},
  \citenamefont {{Goetz}}, \citenamefont {{Hall}}, \citenamefont {{Izumi}},
  \citenamefont {{Kandhasamy}}, \citenamefont {{Karki}}, \citenamefont
  {{Kissel}}, \citenamefont {{Mendell}}, \citenamefont {{Savage}},
  \citenamefont {{Tuyenbayev}}, \citenamefont {{Urban}}, \citenamefont
  {{Viets}}, \citenamefont {{Wade}},\ and\ \citenamefont
  {{Weinstein}}}]{2017PhRvD..96j2001C}%
  \BibitemOpen
  \bibfield  {author} {\bibinfo {author} {\bibfnamefont {C.}~\bibnamefont
  {{Cahillane}}}, \bibinfo {author} {\bibfnamefont {J.}~\bibnamefont
  {{Betzwieser}}}, \bibinfo {author} {\bibfnamefont {D.~A.}\ \bibnamefont
  {{Brown}}}, \bibinfo {author} {\bibfnamefont {E.}~\bibnamefont {{Goetz}}},
  \bibinfo {author} {\bibfnamefont {E.~D.}\ \bibnamefont {{Hall}}}, \bibinfo
  {author} {\bibfnamefont {K.}~\bibnamefont {{Izumi}}}, \bibinfo {author}
  {\bibfnamefont {S.}~\bibnamefont {{Kandhasamy}}}, \bibinfo {author}
  {\bibfnamefont {S.}~\bibnamefont {{Karki}}}, \bibinfo {author} {\bibfnamefont
  {J.~S.}\ \bibnamefont {{Kissel}}}, \bibinfo {author} {\bibfnamefont
  {G.}~\bibnamefont {{Mendell}}}, \bibinfo {author} {\bibfnamefont {R.~L.}\
  \bibnamefont {{Savage}}}, \bibinfo {author} {\bibfnamefont {D.}~\bibnamefont
  {{Tuyenbayev}}}, \bibinfo {author} {\bibfnamefont {A.}~\bibnamefont
  {{Urban}}}, \bibinfo {author} {\bibfnamefont {A.}~\bibnamefont {{Viets}}},
  \bibinfo {author} {\bibfnamefont {M.}~\bibnamefont {{Wade}}}, \ and\ \bibinfo
  {author} {\bibfnamefont {A.~J.}\ \bibnamefont {{Weinstein}}},\ }\href
  {\doibase 10.1103/PhysRevD.96.102001} {\bibfield  {journal} {\bibinfo
  {journal} {\prd}\ }\textbf {\bibinfo {volume} {96}},\ \bibinfo {eid} {102001}
  (\bibinfo {year} {2017})},\ \Eprint {http://arxiv.org/abs/1708.03023}
  {arXiv:1708.03023 [astro-ph.IM]} \BibitemShut {NoStop}%
\bibitem [{\citenamefont {{Moore}}\ and\ \citenamefont
  {{Gair}}(2014)}]{2014PhRvL.113y1101M}%
  \BibitemOpen
  \bibfield  {author} {\bibinfo {author} {\bibfnamefont {C.~J.}\ \bibnamefont
  {{Moore}}}\ and\ \bibinfo {author} {\bibfnamefont {J.~R.}\ \bibnamefont
  {{Gair}}},\ }\href {\doibase 10.1103/PhysRevLett.113.251101} {\bibfield
  {journal} {\bibinfo  {journal} {Physical Review Letters}\ }\textbf {\bibinfo
  {volume} {113}},\ \bibinfo {eid} {251101} (\bibinfo {year} {2014})},\ \Eprint
  {http://arxiv.org/abs/1412.3657} {arXiv:1412.3657 [gr-qc]} \BibitemShut
  {NoStop}%
\bibitem [{\citenamefont {{Moore}}\ \emph {et~al.}(2016)\citenamefont
  {{Moore}}, \citenamefont {{Berry}}, \citenamefont {{Chua}},\ and\
  \citenamefont {{Gair}}}]{2016PhRvD..93f4001M}%
  \BibitemOpen
  \bibfield  {author} {\bibinfo {author} {\bibfnamefont {C.~J.}\ \bibnamefont
  {{Moore}}}, \bibinfo {author} {\bibfnamefont {C.~P.~L.}\ \bibnamefont
  {{Berry}}}, \bibinfo {author} {\bibfnamefont {A.~J.~K.}\ \bibnamefont
  {{Chua}}}, \ and\ \bibinfo {author} {\bibfnamefont {J.~R.}\ \bibnamefont
  {{Gair}}},\ }\href {\doibase 10.1103/PhysRevD.93.064001} {\bibfield
  {journal} {\bibinfo  {journal} {\prd}\ }\textbf {\bibinfo {volume} {93}},\
  \bibinfo {eid} {064001} (\bibinfo {year} {2016})},\ \Eprint
  {http://arxiv.org/abs/1509.04066} {arXiv:1509.04066 [gr-qc]} \BibitemShut
  {NoStop}%
\bibitem [{\citenamefont {{Doctor}}\ \emph {et~al.}(2017)\citenamefont
  {{Doctor}}, \citenamefont {{Farr}}, \citenamefont {{Holz}},\ and\
  \citenamefont {{P{\"u}rrer}}}]{2017PhRvD..96l3011D}%
  \BibitemOpen
  \bibfield  {author} {\bibinfo {author} {\bibfnamefont {Z.}~\bibnamefont
  {{Doctor}}}, \bibinfo {author} {\bibfnamefont {B.}~\bibnamefont {{Farr}}},
  \bibinfo {author} {\bibfnamefont {D.~E.}\ \bibnamefont {{Holz}}}, \ and\
  \bibinfo {author} {\bibfnamefont {M.}~\bibnamefont {{P{\"u}rrer}}},\ }\href
  {\doibase 10.1103/PhysRevD.96.123011} {\bibfield  {journal} {\bibinfo
  {journal} {\prd}\ }\textbf {\bibinfo {volume} {96}},\ \bibinfo {eid} {123011}
  (\bibinfo {year} {2017})},\ \Eprint {http://arxiv.org/abs/1706.05408}
  {arXiv:1706.05408 [astro-ph.HE]} \BibitemShut {NoStop}%
\bibitem [{\citenamefont {{Huerta}}\ \emph {et~al.}(2018)\citenamefont
  {{Huerta}}, \citenamefont {{Moore}}, \citenamefont {{Kumar}}, \citenamefont
  {{George}}, \citenamefont {{Chua}}, \citenamefont {{Haas}}, \citenamefont
  {{Wessel}}, \citenamefont {{Johnson}}, \citenamefont {{Glennon}},
  \citenamefont {{Rebei}}, \citenamefont {{Holgado}}, \citenamefont {{Gair}},\
  and\ \citenamefont {{Pfeiffer}}}]{2018PhRvD..97b4031H}%
  \BibitemOpen
  \bibfield  {author} {\bibinfo {author} {\bibfnamefont {E.~A.}\ \bibnamefont
  {{Huerta}}}, \bibinfo {author} {\bibfnamefont {C.~J.}\ \bibnamefont
  {{Moore}}}, \bibinfo {author} {\bibfnamefont {P.}~\bibnamefont {{Kumar}}},
  \bibinfo {author} {\bibfnamefont {D.}~\bibnamefont {{George}}}, \bibinfo
  {author} {\bibfnamefont {A.~J.~K.}\ \bibnamefont {{Chua}}}, \bibinfo {author}
  {\bibfnamefont {R.}~\bibnamefont {{Haas}}}, \bibinfo {author} {\bibfnamefont
  {E.}~\bibnamefont {{Wessel}}}, \bibinfo {author} {\bibfnamefont
  {D.}~\bibnamefont {{Johnson}}}, \bibinfo {author} {\bibfnamefont
  {D.}~\bibnamefont {{Glennon}}}, \bibinfo {author} {\bibfnamefont
  {A.}~\bibnamefont {{Rebei}}}, \bibinfo {author} {\bibfnamefont {A.~M.}\
  \bibnamefont {{Holgado}}}, \bibinfo {author} {\bibfnamefont {J.~R.}\
  \bibnamefont {{Gair}}}, \ and\ \bibinfo {author} {\bibfnamefont {H.~P.}\
  \bibnamefont {{Pfeiffer}}},\ }\href {\doibase 10.1103/PhysRevD.97.024031}
  {\bibfield  {journal} {\bibinfo  {journal} {\prd}\ }\textbf {\bibinfo
  {volume} {97}},\ \bibinfo {eid} {024031} (\bibinfo {year} {2018})},\ \Eprint
  {http://arxiv.org/abs/1711.06276} {arXiv:1711.06276 [gr-qc]} \BibitemShut
  {NoStop}%
\bibitem [{\citenamefont {{Taylor}}\ and\ \citenamefont
  {{Gerosa}}(2018)}]{2018arXiv180608365T}%
  \BibitemOpen
  \bibfield  {author} {\bibinfo {author} {\bibfnamefont {S.~R.}\ \bibnamefont
  {{Taylor}}}\ and\ \bibinfo {author} {\bibfnamefont {D.}~\bibnamefont
  {{Gerosa}}},\ }\href {\doibase 10.1103/PhysRevD.98.083017} {\bibfield
  {journal} {\bibinfo  {journal} {\prd}\ }\textbf {\bibinfo {volume} {98}},\
  \bibinfo {pages} {083017} (\bibinfo {year} {2018})},\ \Eprint
  {http://arxiv.org/abs/1806.08365} {arXiv:1806.08365 [astro-ph.HE]}
  \BibitemShut {NoStop}%
\bibitem [{\citenamefont {{LIGO Scientific Collaboration}}\ and\ \citenamefont
  {{Virgo Collaboration}}()}]{LAL}%
  \BibitemOpen
  \bibfield  {author} {\bibinfo {author} {\bibnamefont {{LIGO Scientific
  Collaboration}}}\ and\ \bibinfo {author} {\bibnamefont {{Virgo
  Collaboration}}},\ }\href@noop {} {\ }\bibinfo {note}
  {\href{https://git.ligo.org/lscsoft/lalsuite}{git.ligo.org/lscsoft/lalsuite}}\BibitemShut
  {NoStop}%
\bibitem [{\citenamefont {{Cutler}}\ and\ \citenamefont
  {{Flanagan}}(1994)}]{1994PhRvD..49.2658C}%
  \BibitemOpen
  \bibfield  {author} {\bibinfo {author} {\bibfnamefont {C.}~\bibnamefont
  {{Cutler}}}\ and\ \bibinfo {author} {\bibfnamefont {{\'E}.~E.}\ \bibnamefont
  {{Flanagan}}},\ }\href {\doibase 10.1103/PhysRevD.49.2658} {\bibfield
  {journal} {\bibinfo  {journal} {\prd}\ }\textbf {\bibinfo {volume} {49}},\
  \bibinfo {pages} {2658} (\bibinfo {year} {1994})},\ \Eprint
  {http://arxiv.org/abs/gr-qc/9402014} {gr-qc/9402014} \BibitemShut {NoStop}%
\bibitem [{\citenamefont {{Poisson}}\ and\ \citenamefont
  {{Will}}(1995)}]{1995PhRvD..52..848P}%
  \BibitemOpen
  \bibfield  {author} {\bibinfo {author} {\bibfnamefont {E.}~\bibnamefont
  {{Poisson}}}\ and\ \bibinfo {author} {\bibfnamefont {C.~M.}\ \bibnamefont
  {{Will}}},\ }\href {\doibase 10.1103/PhysRevD.52.848} {\bibfield  {journal}
  {\bibinfo  {journal} {\prd}\ }\textbf {\bibinfo {volume} {52}},\ \bibinfo
  {pages} {848} (\bibinfo {year} {1995})},\ \Eprint
  {http://arxiv.org/abs/gr-qc/9502040} {gr-qc/9502040} \BibitemShut {NoStop}%
\bibitem [{\citenamefont {{Mackay}}(2003)}]{2003itil.book.....M}%
  \BibitemOpen
  \bibfield  {author} {\bibinfo {author} {\bibfnamefont {D.~J.~C.}\
  \bibnamefont {{Mackay}}},\ }\href@noop {} {\emph {\bibinfo {title}
  {Information Theory, Inference and Learning Algorithms, by David
  J.~C.~MacKay, pp.~640.~ISBN 0521642981.~Cambridge, UK: Cambridge University
  Press, October 2003.}}}\ (\bibinfo {year} {2003})\ p.\ \bibinfo {pages}
  {640}\BibitemShut {NoStop}%
\bibitem [{\citenamefont {{Ajith}}(2011)}]{2011PhRvD..84h4037A}%
  \BibitemOpen
  \bibfield  {author} {\bibinfo {author} {\bibfnamefont {P.}~\bibnamefont
  {{Ajith}}},\ }\href {\doibase 10.1103/PhysRevD.84.084037} {\bibfield
  {journal} {\bibinfo  {journal} {\prd}\ }\textbf {\bibinfo {volume} {84}},\
  \bibinfo {eid} {084037} (\bibinfo {year} {2011})},\ \Eprint
  {http://arxiv.org/abs/1107.1267} {arXiv:1107.1267 [gr-qc]} \BibitemShut
  {NoStop}%
\bibitem [{\citenamefont {{Pedregosa}}\ \emph {et~al.}(2012)\citenamefont
  {{Pedregosa}}, \citenamefont {{Varoquaux}}, \citenamefont {{Gramfort}},
  \citenamefont {{Michel}}, \citenamefont {{Thirion}}, \citenamefont
  {{Grisel}}, \citenamefont {{Blondel}}, \citenamefont {{M{\"u}ller}},
  \citenamefont {{Nothman}}, \citenamefont {{Louppe}}, \citenamefont
  {{Prettenhofer}}, \citenamefont {{Weiss}}, \citenamefont {{Dubourg}},
  \citenamefont {{Vanderplas}}, \citenamefont {{Passos}}, \citenamefont
  {{Cournapeau}}, \citenamefont {{Brucher}}, \citenamefont {{Perrot}},\ and\
  \citenamefont {{Duchesnay}}}]{2012arXiv1201.0490P}%
  \BibitemOpen
  \bibfield  {author} {\bibinfo {author} {\bibfnamefont {F.}~\bibnamefont
  {{Pedregosa}}}, \bibinfo {author} {\bibfnamefont {G.}~\bibnamefont
  {{Varoquaux}}}, \bibinfo {author} {\bibfnamefont {A.}~\bibnamefont
  {{Gramfort}}}, \bibinfo {author} {\bibfnamefont {V.}~\bibnamefont
  {{Michel}}}, \bibinfo {author} {\bibfnamefont {B.}~\bibnamefont {{Thirion}}},
  \bibinfo {author} {\bibfnamefont {O.}~\bibnamefont {{Grisel}}}, \bibinfo
  {author} {\bibfnamefont {M.}~\bibnamefont {{Blondel}}}, \bibinfo {author}
  {\bibfnamefont {A.}~\bibnamefont {{M{\"u}ller}}}, \bibinfo {author}
  {\bibfnamefont {J.}~\bibnamefont {{Nothman}}}, \bibinfo {author}
  {\bibfnamefont {G.}~\bibnamefont {{Louppe}}}, \bibinfo {author}
  {\bibfnamefont {P.}~\bibnamefont {{Prettenhofer}}}, \bibinfo {author}
  {\bibfnamefont {R.}~\bibnamefont {{Weiss}}}, \bibinfo {author} {\bibfnamefont
  {V.}~\bibnamefont {{Dubourg}}}, \bibinfo {author} {\bibfnamefont
  {J.}~\bibnamefont {{Vanderplas}}}, \bibinfo {author} {\bibfnamefont
  {A.}~\bibnamefont {{Passos}}}, \bibinfo {author} {\bibfnamefont
  {D.}~\bibnamefont {{Cournapeau}}}, \bibinfo {author} {\bibfnamefont
  {M.}~\bibnamefont {{Brucher}}}, \bibinfo {author} {\bibfnamefont
  {M.}~\bibnamefont {{Perrot}}}, \ and\ \bibinfo {author} {\bibfnamefont
  {{\'E}.}~\bibnamefont {{Duchesnay}}},\ }\href@noop {} {\bibfield  {journal}
  {\bibinfo  {journal} {Journal of Machine Learning Research}\ }\textbf
  {\bibinfo {volume} {12}},\ \bibinfo {pages} {2825} (\bibinfo {year}
  {2012})},\ \Eprint {http://arxiv.org/abs/1201.0490} {1201.0490} \BibitemShut
  {NoStop}%
\bibitem [{\citenamefont {{Mrou{\'e}}}\ \emph {et~al.}(2013)\citenamefont
  {{Mrou{\'e}}}, \citenamefont {{Scheel}}, \citenamefont {{Szil{\'a}gyi}},
  \citenamefont {{Pfeiffer}}, \citenamefont {{Boyle}}, \citenamefont
  {{Hemberger}}, \citenamefont {{Kidder}}, \citenamefont {{Lovelace}},
  \citenamefont {{Ossokine}}, \citenamefont {{Taylor}}, \citenamefont
  {{Zengino{\u g}lu}}, \citenamefont {{Buchman}}, \citenamefont {{Chu}},
  \citenamefont {{Foley}}, \citenamefont {{Giesler}}, \citenamefont {{Owen}},\
  and\ \citenamefont {{Teukolsky}}}]{2013PhRvL.111x1104M}%
  \BibitemOpen
  \bibfield  {author} {\bibinfo {author} {\bibfnamefont {A.~H.}\ \bibnamefont
  {{Mrou{\'e}}}}, \bibinfo {author} {\bibfnamefont {M.~A.}\ \bibnamefont
  {{Scheel}}}, \bibinfo {author} {\bibfnamefont {B.}~\bibnamefont
  {{Szil{\'a}gyi}}}, \bibinfo {author} {\bibfnamefont {H.~P.}\ \bibnamefont
  {{Pfeiffer}}}, \bibinfo {author} {\bibfnamefont {M.}~\bibnamefont {{Boyle}}},
  \bibinfo {author} {\bibfnamefont {D.~A.}\ \bibnamefont {{Hemberger}}},
  \bibinfo {author} {\bibfnamefont {L.~E.}\ \bibnamefont {{Kidder}}}, \bibinfo
  {author} {\bibfnamefont {G.}~\bibnamefont {{Lovelace}}}, \bibinfo {author}
  {\bibfnamefont {S.}~\bibnamefont {{Ossokine}}}, \bibinfo {author}
  {\bibfnamefont {N.~W.}\ \bibnamefont {{Taylor}}}, \bibinfo {author}
  {\bibfnamefont {A.}~\bibnamefont {{Zengino{\u g}lu}}}, \bibinfo {author}
  {\bibfnamefont {L.~T.}\ \bibnamefont {{Buchman}}}, \bibinfo {author}
  {\bibfnamefont {T.}~\bibnamefont {{Chu}}}, \bibinfo {author} {\bibfnamefont
  {E.}~\bibnamefont {{Foley}}}, \bibinfo {author} {\bibfnamefont
  {M.}~\bibnamefont {{Giesler}}}, \bibinfo {author} {\bibfnamefont
  {R.}~\bibnamefont {{Owen}}}, \ and\ \bibinfo {author} {\bibfnamefont {S.~A.}\
  \bibnamefont {{Teukolsky}}},\ }\href {\doibase
  10.1103/PhysRevLett.111.241104} {\bibfield  {journal} {\bibinfo  {journal}
  {\prl}\ }\textbf {\bibinfo {volume} {111}},\ \bibinfo {eid} {241104}
  (\bibinfo {year} {2013})},\ \Eprint {http://arxiv.org/abs/1304.6077}
  {arXiv:1304.6077 [gr-qc]} \BibitemShut {NoStop}%
\bibitem [{\citenamefont {{Kumar}}\ \emph {et~al.}(2015)\citenamefont
  {{Kumar}}, \citenamefont {{Barkett}}, \citenamefont {{Bhagwat}},
  \citenamefont {{Afshari}}, \citenamefont {{Brown}}, \citenamefont
  {{Lovelace}}, \citenamefont {{Scheel}},\ and\ \citenamefont
  {{Szil{\'a}gyi}}}]{2015PhRvD..92j2001K}%
  \BibitemOpen
  \bibfield  {author} {\bibinfo {author} {\bibfnamefont {P.}~\bibnamefont
  {{Kumar}}}, \bibinfo {author} {\bibfnamefont {K.}~\bibnamefont {{Barkett}}},
  \bibinfo {author} {\bibfnamefont {S.}~\bibnamefont {{Bhagwat}}}, \bibinfo
  {author} {\bibfnamefont {N.}~\bibnamefont {{Afshari}}}, \bibinfo {author}
  {\bibfnamefont {D.~A.}\ \bibnamefont {{Brown}}}, \bibinfo {author}
  {\bibfnamefont {G.}~\bibnamefont {{Lovelace}}}, \bibinfo {author}
  {\bibfnamefont {M.~A.}\ \bibnamefont {{Scheel}}}, \ and\ \bibinfo {author}
  {\bibfnamefont {B.}~\bibnamefont {{Szil{\'a}gyi}}},\ }\href {\doibase
  10.1103/PhysRevD.92.102001} {\bibfield  {journal} {\bibinfo  {journal}
  {\prd}\ }\textbf {\bibinfo {volume} {92}},\ \bibinfo {eid} {102001} (\bibinfo
  {year} {2015})},\ \Eprint {http://arxiv.org/abs/1507.00103} {arXiv:1507.00103
  [gr-qc]} \BibitemShut {NoStop}%
\bibitem [{\citenamefont {{Blackman}}\ \emph {et~al.}(2015)\citenamefont
  {{Blackman}}, \citenamefont {{Field}}, \citenamefont {{Galley}},
  \citenamefont {{Szil{\'a}gyi}}, \citenamefont {{Scheel}}, \citenamefont
  {{Tiglio}},\ and\ \citenamefont {{Hemberger}}}]{2015PhRvL.115l1102B}%
  \BibitemOpen
  \bibfield  {author} {\bibinfo {author} {\bibfnamefont {J.}~\bibnamefont
  {{Blackman}}}, \bibinfo {author} {\bibfnamefont {S.~E.}\ \bibnamefont
  {{Field}}}, \bibinfo {author} {\bibfnamefont {C.~R.}\ \bibnamefont
  {{Galley}}}, \bibinfo {author} {\bibfnamefont {B.}~\bibnamefont
  {{Szil{\'a}gyi}}}, \bibinfo {author} {\bibfnamefont {M.~A.}\ \bibnamefont
  {{Scheel}}}, \bibinfo {author} {\bibfnamefont {M.}~\bibnamefont {{Tiglio}}},
  \ and\ \bibinfo {author} {\bibfnamefont {D.~A.}\ \bibnamefont
  {{Hemberger}}},\ }\href {\doibase 10.1103/PhysRevLett.115.121102} {\bibfield
  {journal} {\bibinfo  {journal} {\prl}\ }\textbf {\bibinfo {volume} {115}},\
  \bibinfo {eid} {121102} (\bibinfo {year} {2015})},\ \Eprint
  {http://arxiv.org/abs/1502.07758} {arXiv:1502.07758 [gr-qc]} \BibitemShut
  {NoStop}%
\bibitem [{\citenamefont {{Chu}}\ \emph {et~al.}(2016)\citenamefont {{Chu}},
  \citenamefont {{Fong}}, \citenamefont {{Kumar}}, \citenamefont {{Pfeiffer}},
  \citenamefont {{Boyle}}, \citenamefont {{Hemberger}}, \citenamefont
  {{Kidder}}, \citenamefont {{Scheel}},\ and\ \citenamefont
  {{Szilagyi}}}]{2016CQGra..33p5001C}%
  \BibitemOpen
  \bibfield  {author} {\bibinfo {author} {\bibfnamefont {T.}~\bibnamefont
  {{Chu}}}, \bibinfo {author} {\bibfnamefont {H.}~\bibnamefont {{Fong}}},
  \bibinfo {author} {\bibfnamefont {P.}~\bibnamefont {{Kumar}}}, \bibinfo
  {author} {\bibfnamefont {H.~P.}\ \bibnamefont {{Pfeiffer}}}, \bibinfo
  {author} {\bibfnamefont {M.}~\bibnamefont {{Boyle}}}, \bibinfo {author}
  {\bibfnamefont {D.~A.}\ \bibnamefont {{Hemberger}}}, \bibinfo {author}
  {\bibfnamefont {L.~E.}\ \bibnamefont {{Kidder}}}, \bibinfo {author}
  {\bibfnamefont {M.~A.}\ \bibnamefont {{Scheel}}}, \ and\ \bibinfo {author}
  {\bibfnamefont {B.}~\bibnamefont {{Szilagyi}}},\ }\href {\doibase
  10.1088/0264-9381/33/16/165001} {\bibfield  {journal} {\bibinfo  {journal}
  {\cqg}\ }\textbf {\bibinfo {volume} {33}},\ \bibinfo {eid} {165001} (\bibinfo
  {year} {2016})},\ \Eprint {http://arxiv.org/abs/1512.06800} {arXiv:1512.06800
  [gr-qc]} \BibitemShut {NoStop}%
\bibitem [{\citenamefont {Jones}\ \emph {et~al.}(01  )\citenamefont {Jones},
  \citenamefont {Oliphant}, \citenamefont {Peterson} \emph
  {et~al.}}]{Jones:2001aa}%
  \BibitemOpen
  \bibfield  {author} {\bibinfo {author} {\bibfnamefont {E.}~\bibnamefont
  {Jones}}, \bibinfo {author} {\bibfnamefont {T.}~\bibnamefont {Oliphant}},
  \bibinfo {author} {\bibfnamefont {P.}~\bibnamefont {Peterson}},  \emph
  {et~al.},\ }\href@noop {} {\enquote {\bibinfo {title} {{SciPy}: Open source
  scientific tools for {Python}},}\ }\bibinfo {howpublished}
  {\url{http://www.scipy.org/}} (\bibinfo {year} {2001--})\BibitemShut
  {NoStop}%
\bibitem [{\citenamefont {van~der Walt}\ \emph {et~al.}(2011)\citenamefont
  {van~der Walt}, \citenamefont {Colbert},\ and\ \citenamefont
  {Varoquaux}}]{Walt}%
  \BibitemOpen
  \bibfield  {author} {\bibinfo {author} {\bibfnamefont {S.}~\bibnamefont
  {van~der Walt}}, \bibinfo {author} {\bibfnamefont {S.}~\bibnamefont
  {Colbert}}, \ and\ \bibinfo {author} {\bibfnamefont {G.}~\bibnamefont
  {Varoquaux}},\ }\href {\doibase 10.1109/MCSE.2011.37} {\bibfield  {journal}
  {\bibinfo  {journal} {Computing in Science Engineering}\ }\textbf {\bibinfo
  {volume} {13}},\ \bibinfo {pages} {22} (\bibinfo {year} {2011})}\BibitemShut
  {NoStop}%
\bibitem [{\citenamefont {Collette}(2013)}]{h5py}%
  \BibitemOpen
  \bibfield  {author} {\bibinfo {author} {\bibfnamefont {A.}~\bibnamefont
  {Collette}},\ }\href@noop {} {\emph {\bibinfo {title} {Python and HDF5}}}\
  (\bibinfo  {publisher} {O'Reilly},\ \bibinfo {year} {2013})\BibitemShut
  {NoStop}%
\bibitem [{\citenamefont {{Travis Continuous Integration}}()}]{travis-ci}%
  \BibitemOpen
  \bibfield  {author} {\bibinfo {author} {\bibnamefont {{Travis Continuous
  Integration}}},\ }\href@noop {} {\ }\bibinfo {note}
  {\href{https://travis-ci.org/}{travis-ci.org}}\BibitemShut {NoStop}%
\end{thebibliography}

\begin{thebibliography}{19}%
\makeatletter
\providecommand \@ifxundefined [1]{%
 \@ifx{#1\undefined}
}%
\providecommand \@ifnum [1]{%
 \ifnum #1\expandafter \@firstoftwo
 \else \expandafter \@secondoftwo
 \fi
}%
\providecommand \@ifx [1]{%
 \ifx #1\expandafter \@firstoftwo
 \else \expandafter \@secondoftwo
 \fi
}%
\providecommand \natexlab [1]{#1}%
\providecommand \enquote  [1]{``#1''}%
\providecommand \bibnamefont  [1]{#1}%
\providecommand \bibfnamefont [1]{#1}%
\providecommand \citenamefont [1]{#1}%
\providecommand \href@noop [0]{\@secondoftwo}%
\providecommand \href [0]{\begingroup \@sanitize@url \@href}%
\providecommand \@href[1]{\@@startlink{#1}\@@href}%
\providecommand \@@href[1]{\endgroup#1\@@endlink}%
\providecommand \@sanitize@url [0]{\catcode `\\12\catcode `\$12\catcode
  `\&12\catcode `\#12\catcode `\^12\catcode `\_12\catcode `\%12\relax}%
\providecommand \@@startlink[1]{}%
\providecommand \@@endlink[0]{}%
\providecommand \url  [0]{\begingroup\@sanitize@url \@url }%
\providecommand \@url [1]{\endgroup\@href {#1}{\urlprefix }}%
\providecommand \urlprefix  [0]{URL }%
\providecommand \Eprint [0]{\href }%
\providecommand \doibase [0]{http://dx.doi.org/}%
\providecommand \selectlanguage [0]{\@gobble}%
\providecommand \bibinfo  [0]{\@secondoftwo}%
\providecommand \bibfield  [0]{\@secondoftwo}%
\providecommand \translation [1]{[#1]}%
\providecommand \BibitemOpen [0]{}%
\providecommand \bibitemStop [0]{}%
\providecommand \bibitemNoStop [0]{.\EOS\space}%
\providecommand \EOS [0]{\spacefactor3000\relax}%
\providecommand \BibitemShut  [1]{\csname bibitem#1\endcsname}%
\let\auto@bib@innerbib\@empty
\bibitem [{\citenamefont {{Rasmussen}}\ and\ \citenamefont
  {{Williams}}(2006)}]{supp_2006gpml.book.....R}%
  \BibitemOpen
  \bibfield  {author} {\bibinfo {author} {\bibfnamefont {C.~E.}\ \bibnamefont
  {{Rasmussen}}}\ and\ \bibinfo {author} {\bibfnamefont {C.~K.~I.}\
  \bibnamefont {{Williams}}},\ }\href@noop {} {\emph {\bibinfo {title}
  {Gaussian Processes for Machine Learning, by C.E.~Rasmussen and
  C.K.I.~Williams.~ISBN-13 978-0-262-18253-9}}}\ (\bibinfo {year}
  {2006})\BibitemShut {NoStop}%
\bibitem [{\citenamefont {{Mackay}}(2003)}]{supp_2003itil.book.....M}%
  \BibitemOpen
  \bibfield  {author} {\bibinfo {author} {\bibfnamefont {D.~J.~C.}\
  \bibnamefont {{Mackay}}},\ }\href@noop {} {\emph {\bibinfo {title}
  {Information Theory, Inference and Learning Algorithms, by David
  J.~C.~MacKay, pp.~640.~ISBN 0521642981.~Cambridge, UK: Cambridge University
  Press, October 2003.}}}\ (\bibinfo {year} {2003})\ p.\ \bibinfo {pages}
  {640}\BibitemShut {NoStop}%
\bibitem [{\citenamefont {{Pedregosa}}\ \emph {et~al.}(2012)\citenamefont
  {{Pedregosa}}, \citenamefont {{Varoquaux}}, \citenamefont {{Gramfort}},
  \citenamefont {{Michel}}, \citenamefont {{Thirion}}, \citenamefont
  {{Grisel}}, \citenamefont {{Blondel}}, \citenamefont {{M{\"u}ller}},
  \citenamefont {{Nothman}}, \citenamefont {{Louppe}}, \citenamefont
  {{Prettenhofer}}, \citenamefont {{Weiss}}, \citenamefont {{Dubourg}},
  \citenamefont {{Vanderplas}}, \citenamefont {{Passos}}, \citenamefont
  {{Cournapeau}}, \citenamefont {{Brucher}}, \citenamefont {{Perrot}},\ and\
  \citenamefont {{Duchesnay}}}]{supp_2012arXiv1201.0490P}%
  \BibitemOpen
  \bibfield  {author} {\bibinfo {author} {\bibfnamefont {F.}~\bibnamefont
  {{Pedregosa}}}, \bibinfo {author} {\bibfnamefont {G.}~\bibnamefont
  {{Varoquaux}}}, \bibinfo {author} {\bibfnamefont {A.}~\bibnamefont
  {{Gramfort}}}, \bibinfo {author} {\bibfnamefont {V.}~\bibnamefont
  {{Michel}}}, \bibinfo {author} {\bibfnamefont {B.}~\bibnamefont {{Thirion}}},
  \bibinfo {author} {\bibfnamefont {O.}~\bibnamefont {{Grisel}}}, \bibinfo
  {author} {\bibfnamefont {M.}~\bibnamefont {{Blondel}}}, \bibinfo {author}
  {\bibfnamefont {A.}~\bibnamefont {{M{\"u}ller}}}, \bibinfo {author}
  {\bibfnamefont {J.}~\bibnamefont {{Nothman}}}, \bibinfo {author}
  {\bibfnamefont {G.}~\bibnamefont {{Louppe}}}, \bibinfo {author}
  {\bibfnamefont {P.}~\bibnamefont {{Prettenhofer}}}, \bibinfo {author}
  {\bibfnamefont {R.}~\bibnamefont {{Weiss}}}, \bibinfo {author} {\bibfnamefont
  {V.}~\bibnamefont {{Dubourg}}}, \bibinfo {author} {\bibfnamefont
  {J.}~\bibnamefont {{Vanderplas}}}, \bibinfo {author} {\bibfnamefont
  {A.}~\bibnamefont {{Passos}}}, \bibinfo {author} {\bibfnamefont
  {D.}~\bibnamefont {{Cournapeau}}}, \bibinfo {author} {\bibfnamefont
  {M.}~\bibnamefont {{Brucher}}}, \bibinfo {author} {\bibfnamefont
  {M.}~\bibnamefont {{Perrot}}}, \ and\ \bibinfo {author} {\bibfnamefont
  {{\'E}.}~\bibnamefont {{Duchesnay}}},\ }\href@noop {} {\bibfield  {journal}
  {\bibinfo  {journal} {Journal of Machine Learning Research}\ }\textbf
  {\bibinfo {volume} {12}},\ \bibinfo {pages} {2825} (\bibinfo {year}
  {2012})},\ \Eprint {http://arxiv.org/abs/1201.0490} {1201.0490} \BibitemShut
  {NoStop}%
\bibitem [{\citenamefont {{Khan}}\ \emph {et~al.}(2016)\citenamefont {{Khan}},
  \citenamefont {{Husa}}, \citenamefont {{Hannam}}, \citenamefont {{Ohme}},
  \citenamefont {{P{\"u}rrer}}, \citenamefont {{Forteza}},\ and\ \citenamefont
  {{Boh{\'e}}}}]{supp_2016PhRvD..93d4007K}%
  \BibitemOpen
  \bibfield  {author} {\bibinfo {author} {\bibfnamefont {S.}~\bibnamefont
  {{Khan}}}, \bibinfo {author} {\bibfnamefont {S.}~\bibnamefont {{Husa}}},
  \bibinfo {author} {\bibfnamefont {M.}~\bibnamefont {{Hannam}}}, \bibinfo
  {author} {\bibfnamefont {F.}~\bibnamefont {{Ohme}}}, \bibinfo {author}
  {\bibfnamefont {M.}~\bibnamefont {{P{\"u}rrer}}}, \bibinfo {author}
  {\bibfnamefont {X.~J.}\ \bibnamefont {{Forteza}}}, \ and\ \bibinfo {author}
  {\bibfnamefont {A.}~\bibnamefont {{Boh{\'e}}}},\ }\href {\doibase
  10.1103/PhysRevD.93.044007} {\bibfield  {journal} {\bibinfo  {journal}
  {\prd}\ }\textbf {\bibinfo {volume} {93}},\ \bibinfo {eid} {044007} (\bibinfo
  {year} {2016})},\ \Eprint {http://arxiv.org/abs/1508.07253} {arXiv:1508.07253
  [gr-qc]} \BibitemShut {NoStop}%
\bibitem [{\citenamefont {{Ajith}}(2011)}]{supp_2011PhRvD..84h4037A}%
  \BibitemOpen
  \bibfield  {author} {\bibinfo {author} {\bibfnamefont {P.}~\bibnamefont
  {{Ajith}}},\ }\href {\doibase 10.1103/PhysRevD.84.084037} {\bibfield
  {journal} {\bibinfo  {journal} {\prd}\ }\textbf {\bibinfo {volume} {84}},\
  \bibinfo {eid} {084037} (\bibinfo {year} {2011})},\ \Eprint
  {http://arxiv.org/abs/1107.1267} {arXiv:1107.1267 [gr-qc]} \BibitemShut
  {NoStop}%
\bibitem [{\citenamefont {{Cutler}}\ and\ \citenamefont
  {{Flanagan}}(1994)}]{supp_1994PhRvD..49.2658C}%
  \BibitemOpen
  \bibfield  {author} {\bibinfo {author} {\bibfnamefont {C.}~\bibnamefont
  {{Cutler}}}\ and\ \bibinfo {author} {\bibfnamefont {{\'E}.~E.}\ \bibnamefont
  {{Flanagan}}},\ }\href {\doibase 10.1103/PhysRevD.49.2658} {\bibfield
  {journal} {\bibinfo  {journal} {\prd}\ }\textbf {\bibinfo {volume} {49}},\
  \bibinfo {pages} {2658} (\bibinfo {year} {1994})},\ \Eprint
  {http://arxiv.org/abs/gr-qc/9402014} {gr-qc/9402014} \BibitemShut {NoStop}%
\bibitem [{\citenamefont {{Poisson}}\ and\ \citenamefont
  {{Will}}(1995)}]{supp_1995PhRvD..52..848P}%
  \BibitemOpen
  \bibfield  {author} {\bibinfo {author} {\bibfnamefont {E.}~\bibnamefont
  {{Poisson}}}\ and\ \bibinfo {author} {\bibfnamefont {C.~M.}\ \bibnamefont
  {{Will}}},\ }\href {\doibase 10.1103/PhysRevD.52.848} {\bibfield  {journal}
  {\bibinfo  {journal} {\prd}\ }\textbf {\bibinfo {volume} {52}},\ \bibinfo
  {pages} {848} (\bibinfo {year} {1995})},\ \Eprint
  {http://arxiv.org/abs/gr-qc/9502040} {gr-qc/9502040} \BibitemShut {NoStop}%
\bibitem [{\citenamefont {{Blackman}}\ \emph {et~al.}(2017)\citenamefont
  {{Blackman}}, \citenamefont {{Field}}, \citenamefont {{Scheel}},
  \citenamefont {{Galley}}, \citenamefont {{Ott}}, \citenamefont {{Boyle}},
  \citenamefont {{Kidder}}, \citenamefont {{Pfeiffer}},\ and\ \citenamefont
  {{Szil{\'a}gyi}}}]{supp_2017PhRvD..96b4058B}%
  \BibitemOpen
  \bibfield  {author} {\bibinfo {author} {\bibfnamefont {J.}~\bibnamefont
  {{Blackman}}}, \bibinfo {author} {\bibfnamefont {S.~E.}\ \bibnamefont
  {{Field}}}, \bibinfo {author} {\bibfnamefont {M.~A.}\ \bibnamefont
  {{Scheel}}}, \bibinfo {author} {\bibfnamefont {C.~R.}\ \bibnamefont
  {{Galley}}}, \bibinfo {author} {\bibfnamefont {C.~D.}\ \bibnamefont {{Ott}}},
  \bibinfo {author} {\bibfnamefont {M.}~\bibnamefont {{Boyle}}}, \bibinfo
  {author} {\bibfnamefont {L.~E.}\ \bibnamefont {{Kidder}}}, \bibinfo {author}
  {\bibfnamefont {H.~P.}\ \bibnamefont {{Pfeiffer}}}, \ and\ \bibinfo {author}
  {\bibfnamefont {B.}~\bibnamefont {{Szil{\'a}gyi}}},\ }\href {\doibase
  10.1103/PhysRevD.96.024058} {\bibfield  {journal} {\bibinfo  {journal}
  {\prd}\ }\textbf {\bibinfo {volume} {96}},\ \bibinfo {eid} {024058} (\bibinfo
  {year} {2017})},\ \Eprint {http://arxiv.org/abs/1705.07089} {arXiv:1705.07089
  [gr-qc]} \BibitemShut {NoStop}%
\bibitem [{\citenamefont {{Mrou{\'e}}}\ \emph {et~al.}(2013)\citenamefont
  {{Mrou{\'e}}}, \citenamefont {{Scheel}}, \citenamefont {{Szil{\'a}gyi}},
  \citenamefont {{Pfeiffer}}, \citenamefont {{Boyle}}, \citenamefont
  {{Hemberger}}, \citenamefont {{Kidder}}, \citenamefont {{Lovelace}},
  \citenamefont {{Ossokine}}, \citenamefont {{Taylor}}, \citenamefont
  {{Zengino{\u g}lu}}, \citenamefont {{Buchman}}, \citenamefont {{Chu}},
  \citenamefont {{Foley}}, \citenamefont {{Giesler}}, \citenamefont {{Owen}},\
  and\ \citenamefont {{Teukolsky}}}]{supp_2013PhRvL.111x1104M}%
  \BibitemOpen
  \bibfield  {author} {\bibinfo {author} {\bibfnamefont {A.~H.}\ \bibnamefont
  {{Mrou{\'e}}}}, \bibinfo {author} {\bibfnamefont {M.~A.}\ \bibnamefont
  {{Scheel}}}, \bibinfo {author} {\bibfnamefont {B.}~\bibnamefont
  {{Szil{\'a}gyi}}}, \bibinfo {author} {\bibfnamefont {H.~P.}\ \bibnamefont
  {{Pfeiffer}}}, \bibinfo {author} {\bibfnamefont {M.}~\bibnamefont {{Boyle}}},
  \bibinfo {author} {\bibfnamefont {D.~A.}\ \bibnamefont {{Hemberger}}},
  \bibinfo {author} {\bibfnamefont {L.~E.}\ \bibnamefont {{Kidder}}}, \bibinfo
  {author} {\bibfnamefont {G.}~\bibnamefont {{Lovelace}}}, \bibinfo {author}
  {\bibfnamefont {S.}~\bibnamefont {{Ossokine}}}, \bibinfo {author}
  {\bibfnamefont {N.~W.}\ \bibnamefont {{Taylor}}}, \bibinfo {author}
  {\bibfnamefont {A.}~\bibnamefont {{Zengino{\u g}lu}}}, \bibinfo {author}
  {\bibfnamefont {L.~T.}\ \bibnamefont {{Buchman}}}, \bibinfo {author}
  {\bibfnamefont {T.}~\bibnamefont {{Chu}}}, \bibinfo {author} {\bibfnamefont
  {E.}~\bibnamefont {{Foley}}}, \bibinfo {author} {\bibfnamefont
  {M.}~\bibnamefont {{Giesler}}}, \bibinfo {author} {\bibfnamefont
  {R.}~\bibnamefont {{Owen}}}, \ and\ \bibinfo {author} {\bibfnamefont {S.~A.}\
  \bibnamefont {{Teukolsky}}},\ }\href {\doibase
  10.1103/PhysRevLett.111.241104} {\bibfield  {journal} {\bibinfo  {journal}
  {\prl}\ }\textbf {\bibinfo {volume} {111}},\ \bibinfo {eid} {241104}
  (\bibinfo {year} {2013})},\ \Eprint {http://arxiv.org/abs/1304.6077}
  {arXiv:1304.6077 [gr-qc]} \BibitemShut {NoStop}%
\bibitem [{\citenamefont {{Kumar}}\ \emph {et~al.}(2015)\citenamefont
  {{Kumar}}, \citenamefont {{Barkett}}, \citenamefont {{Bhagwat}},
  \citenamefont {{Afshari}}, \citenamefont {{Brown}}, \citenamefont
  {{Lovelace}}, \citenamefont {{Scheel}},\ and\ \citenamefont
  {{Szil{\'a}gyi}}}]{supp_2015PhRvD..92j2001K}%
  \BibitemOpen
  \bibfield  {author} {\bibinfo {author} {\bibfnamefont {P.}~\bibnamefont
  {{Kumar}}}, \bibinfo {author} {\bibfnamefont {K.}~\bibnamefont {{Barkett}}},
  \bibinfo {author} {\bibfnamefont {S.}~\bibnamefont {{Bhagwat}}}, \bibinfo
  {author} {\bibfnamefont {N.}~\bibnamefont {{Afshari}}}, \bibinfo {author}
  {\bibfnamefont {D.~A.}\ \bibnamefont {{Brown}}}, \bibinfo {author}
  {\bibfnamefont {G.}~\bibnamefont {{Lovelace}}}, \bibinfo {author}
  {\bibfnamefont {M.~A.}\ \bibnamefont {{Scheel}}}, \ and\ \bibinfo {author}
  {\bibfnamefont {B.}~\bibnamefont {{Szil{\'a}gyi}}},\ }\href {\doibase
  10.1103/PhysRevD.92.102001} {\bibfield  {journal} {\bibinfo  {journal}
  {\prd}\ }\textbf {\bibinfo {volume} {92}},\ \bibinfo {eid} {102001} (\bibinfo
  {year} {2015})},\ \Eprint {http://arxiv.org/abs/1507.00103} {arXiv:1507.00103
  [gr-qc]} \BibitemShut {NoStop}%
\bibitem [{\citenamefont {{Blackman}}\ \emph {et~al.}(2015)\citenamefont
  {{Blackman}}, \citenamefont {{Field}}, \citenamefont {{Galley}},
  \citenamefont {{Szil{\'a}gyi}}, \citenamefont {{Scheel}}, \citenamefont
  {{Tiglio}},\ and\ \citenamefont {{Hemberger}}}]{supp_2015PhRvL.115l1102B}%
  \BibitemOpen
  \bibfield  {author} {\bibinfo {author} {\bibfnamefont {J.}~\bibnamefont
  {{Blackman}}}, \bibinfo {author} {\bibfnamefont {S.~E.}\ \bibnamefont
  {{Field}}}, \bibinfo {author} {\bibfnamefont {C.~R.}\ \bibnamefont
  {{Galley}}}, \bibinfo {author} {\bibfnamefont {B.}~\bibnamefont
  {{Szil{\'a}gyi}}}, \bibinfo {author} {\bibfnamefont {M.~A.}\ \bibnamefont
  {{Scheel}}}, \bibinfo {author} {\bibfnamefont {M.}~\bibnamefont {{Tiglio}}},
  \ and\ \bibinfo {author} {\bibfnamefont {D.~A.}\ \bibnamefont
  {{Hemberger}}},\ }\href {\doibase 10.1103/PhysRevLett.115.121102} {\bibfield
  {journal} {\bibinfo  {journal} {\prl}\ }\textbf {\bibinfo {volume} {115}},\
  \bibinfo {eid} {121102} (\bibinfo {year} {2015})},\ \Eprint
  {http://arxiv.org/abs/1502.07758} {arXiv:1502.07758 [gr-qc]} \BibitemShut
  {NoStop}%
\bibitem [{\citenamefont {{Chu}}\ \emph {et~al.}(2016)\citenamefont {{Chu}},
  \citenamefont {{Fong}}, \citenamefont {{Kumar}}, \citenamefont {{Pfeiffer}},
  \citenamefont {{Boyle}}, \citenamefont {{Hemberger}}, \citenamefont
  {{Kidder}}, \citenamefont {{Scheel}},\ and\ \citenamefont
  {{Szilagyi}}}]{supp_2016CQGra..33p5001C}%
  \BibitemOpen
  \bibfield  {author} {\bibinfo {author} {\bibfnamefont {T.}~\bibnamefont
  {{Chu}}}, \bibinfo {author} {\bibfnamefont {H.}~\bibnamefont {{Fong}}},
  \bibinfo {author} {\bibfnamefont {P.}~\bibnamefont {{Kumar}}}, \bibinfo
  {author} {\bibfnamefont {H.~P.}\ \bibnamefont {{Pfeiffer}}}, \bibinfo
  {author} {\bibfnamefont {M.}~\bibnamefont {{Boyle}}}, \bibinfo {author}
  {\bibfnamefont {D.~A.}\ \bibnamefont {{Hemberger}}}, \bibinfo {author}
  {\bibfnamefont {L.~E.}\ \bibnamefont {{Kidder}}}, \bibinfo {author}
  {\bibfnamefont {M.~A.}\ \bibnamefont {{Scheel}}}, \ and\ \bibinfo {author}
  {\bibfnamefont {B.}~\bibnamefont {{Szilagyi}}},\ }\href {\doibase
  10.1088/0264-9381/33/16/165001} {\bibfield  {journal} {\bibinfo  {journal}
  {\cqg}\ }\textbf {\bibinfo {volume} {33}},\ \bibinfo {eid} {165001} (\bibinfo
  {year} {2016})},\ \Eprint {http://arxiv.org/abs/1512.06800} {arXiv:1512.06800
  [gr-qc]} \BibitemShut {NoStop}%
\bibitem [{\citenamefont {{Boyle}}\ \emph {et~al.}(2018)\citenamefont {{Boyle}}
  \emph {et~al.}}]{supp_Catalog2018}%
  \BibitemOpen
  \bibfield  {author} {\bibinfo {author} {\bibfnamefont {M.}~\bibnamefont
  {{Boyle}}} \emph {et~al.},\ }\href@noop {} {\  (\bibinfo {year} {2018})},\
  \bibinfo {note} {in preparation}\BibitemShut {NoStop}%
\bibitem [{\citenamefont {{Varma}}\ \emph {et~al.}()\citenamefont {{Varma}}
  \emph {et~al.}}]{supp_surfinBH}%
  \BibitemOpen
  \bibfield  {author} {\bibinfo {author} {\bibfnamefont {V.}~\bibnamefont
  {{Varma}}} \emph {et~al.},\ }\href@noop {} {\ }\bibinfo {note}
  {\href{https://pypi.org/project/surfinBH}{pypi.org/project/surfinBH},
  \href{https://doi.org/10.5281/zenodo.1418525}{doi.org/10.5281/zenodo.1418525}}\BibitemShut
  {NoStop}%
\bibitem [{\citenamefont {van~der Walt}\ \emph {et~al.}(2011)\citenamefont
  {van~der Walt}, \citenamefont {Colbert},\ and\ \citenamefont
  {Varoquaux}}]{supp_Walt}%
  \BibitemOpen
  \bibfield  {author} {\bibinfo {author} {\bibfnamefont {S.}~\bibnamefont
  {van~der Walt}}, \bibinfo {author} {\bibfnamefont {S.}~\bibnamefont
  {Colbert}}, \ and\ \bibinfo {author} {\bibfnamefont {G.}~\bibnamefont
  {Varoquaux}},\ }\href {\doibase 10.1109/MCSE.2011.37} {\bibfield  {journal}
  {\bibinfo  {journal} {Computing in Science Engineering}\ }\textbf {\bibinfo
  {volume} {13}},\ \bibinfo {pages} {22} (\bibinfo {year} {2011})}\BibitemShut
  {NoStop}%
\bibitem [{\citenamefont {Jones}\ \emph {et~al.}(01  )\citenamefont {Jones},
  \citenamefont {Oliphant}, \citenamefont {Peterson} \emph
  {et~al.}}]{supp_Jones:2001aa}%
  \BibitemOpen
  \bibfield  {author} {\bibinfo {author} {\bibfnamefont {E.}~\bibnamefont
  {Jones}}, \bibinfo {author} {\bibfnamefont {T.}~\bibnamefont {Oliphant}},
  \bibinfo {author} {\bibfnamefont {P.}~\bibnamefont {Peterson}},  \emph
  {et~al.},\ }\href@noop {} {\enquote {\bibinfo {title} {{SciPy}: Open source
  scientific tools for {Python}},}\ }\bibinfo {howpublished}
  {\url{http://www.scipy.org/}} (\bibinfo {year} {2001--})\BibitemShut
  {NoStop}%
\bibitem [{\citenamefont {Collette}(2013)}]{supp_h5py}%
  \BibitemOpen
  \bibfield  {author} {\bibinfo {author} {\bibfnamefont {A.}~\bibnamefont
  {Collette}},\ }\href@noop {} {\emph {\bibinfo {title} {Python and HDF5}}}\
  (\bibinfo  {publisher} {O'Reilly},\ \bibinfo {year} {2013})\BibitemShut
  {NoStop}%
\bibitem [{\citenamefont {{LIGO Scientific Collaboration}}\ and\ \citenamefont
  {{Virgo Collaboration}}()}]{supp_LAL}%
  \BibitemOpen
  \bibfield  {author} {\bibinfo {author} {\bibnamefont {{LIGO Scientific
  Collaboration}}}\ and\ \bibinfo {author} {\bibnamefont {{Virgo
  Collaboration}}},\ }\href@noop {} {\ }\bibinfo {note}
  {\href{https://git.ligo.org/lscsoft/lalsuite}{git.ligo.org/lscsoft/lalsuite}}\BibitemShut
  {NoStop}%
\bibitem [{\citenamefont {{Travis Continuous Integration}}()}]{supp_travis-ci}%
  \BibitemOpen
  \bibfield  {author} {\bibinfo {author} {\bibnamefont {{Travis Continuous
  Integration}}},\ }\href@noop {} {\ }\bibinfo {note}
  {\href{https://travis-ci.org/}{travis-ci.org}}\BibitemShut {NoStop}%
\end{thebibliography}
